\documentclass[aps,pra,twocolumn,superscriptaddress]{revtex4-1}

\usepackage{graphicx}
\usepackage{amsmath,amssymb,mathtools,braket}
\usepackage{hyperref}
\DeclarePairedDelimiter\floor{\lfloor}{\rfloor}
\usepackage[margin=1.64cm, bmargin=2.25cm]{geometry}

\newcommand{\ad}{\hat{a}}
\newcommand{\au}{\hat{a}^\dagger}
\newcommand{\bd}{\hat{b}}
\newcommand{\bu}{\hat{b}^\dagger}

\usepackage{xcolor}
\hypersetup{
    colorlinks=true,
    linkcolor={blue!100!black},
    citecolor={blue!100!black},
    urlcolor={blue!100!black}
}

\begin{document}
\title{Photon-number entanglement generated by sequential excitation of a two-level atom}

\author{Stephen C. Wein}
	\email{wein.stephen@gmail.com}
	\affiliation{Institute for Quantum Science and Technology and Department of Physics and Astronomy, University of Calgary, Calgary, Canada T2N 1N4}
	\affiliation{Universit\'{e} Grenoble Alpes, CNRS, Grenoble INP, Institut N\'eel, 38000 Grenoble, France}
	
\author{Juan C. Loredo}
	\affiliation{Centre for Nanosciences and Nanotechnology, CNRS, Universit\'e Paris-Saclay, UMR 9001, 10 Boulevard Thomas Gobert, 91120 Palaiseau, France}
	\affiliation{University of Vienna, Faculty of Physics, Vienna Center for Quantum Science and Technology (VCQ), Vienna, Austria}
    \affiliation{Christian Doppler Laboratory for Photonic Quantum Computer, Faculty of Physics, University of Vienna, 1090 Vienna, Austria}

\author{Maria Maffei}
	\affiliation{Universit\'{e} Grenoble Alpes, CNRS, Grenoble INP, Institut N\'eel, 38000 Grenoble, France}

\author{Paul Hilaire}
	\affiliation{Centre for Nanosciences and Nanotechnology, CNRS, Universit\'e Paris-Saclay, UMR 9001, 10 Boulevard Thomas Gobert, 91120 Palaiseau, France}
	\affiliation{Department of Physics, Virginia Tech, Blacksburg, Virginia 24061, USA}
	
\author{Abdelmounaim Harouri}
	\affiliation{Centre for Nanosciences and Nanotechnology, CNRS, Universit\'e Paris-Saclay, UMR 9001, 10 Boulevard Thomas Gobert, 91120 Palaiseau, France}	
	
\author{Niccolo Somaschi}
	\affiliation{Quandela SAS, 10 Boulevard Thomas Gobert, 91120 Palaiseau, France}	
	
\author{Aristide Lema\^itre}
	\affiliation{Centre for Nanosciences and Nanotechnology, CNRS, Universit\'e Paris-Saclay, UMR 9001, 10 Boulevard Thomas Gobert, 91120 Palaiseau, France}

\author{Isabelle Sagnes}
	\affiliation{Centre for Nanosciences and Nanotechnology, CNRS, Universit\'e Paris-Saclay, UMR 9001, 10 Boulevard Thomas Gobert, 91120 Palaiseau, France}

\author{Lo\"{i}c Lanco}
	\affiliation{Centre for Nanosciences and Nanotechnology, CNRS, Universit\'e Paris-Saclay, UMR 9001, 10 Boulevard Thomas Gobert, 91120 Palaiseau, France}	
	\affiliation{Universit\'e de Paris, Centre for Nanoscience and Nanotechnology (C2N), F-91120 Palaiseau, France}
	
\author{Olivier Krebs}
	\affiliation{Centre for Nanosciences and Nanotechnology, CNRS, Universit\'e Paris-Saclay, UMR 9001, 10 Boulevard Thomas Gobert, 91120 Palaiseau, France}
	
\author{Alexia Auff\`eves}
	\affiliation{Universit\'{e} Grenoble Alpes, CNRS, Grenoble INP, Institut N\'eel, 38000 Grenoble, France}

\author{Christoph Simon}
	\affiliation{Institute for Quantum Science and Technology and Department of Physics and Astronomy, University of Calgary, Calgary, Canada T2N 1N4}

\author{Pascale Senellart}
	\affiliation{Centre for Nanosciences and Nanotechnology, CNRS, Universit\'e Paris-Saclay, UMR 9001, 10 Boulevard Thomas Gobert, 91120 Palaiseau, France}

\author{Carlos Ant\'on-Solanas}
    \email{c.anton.so@gmail.com}
	\affiliation{Centre for Nanosciences and Nanotechnology, CNRS, Universit\'e Paris-Saclay, UMR 9001, 10 Boulevard Thomas Gobert, 91120 Palaiseau, France}
	\affiliation{Institute of Physics, Carl von Ossietzky University, 26129 Oldenburg, Germany}
	
\begin{abstract}{
Entanglement and spontaneous emission are fundamental quantum phenomena that drive many applications of quantum physics. During the spontaneous emission of light from an excited two-level atom, the atom briefly becomes entangled with the photonic field. Here, we show that this natural process can be used to produce photon-number entangled states of light distributed in time. By exciting a quantum dot --an artificial two-level atom-- with two sequential $\pi$ pulses, we generate a photon-number Bell state. We characterise this state using time-resolved intensity and phase correlation measurements. Furthermore, we theoretically show that applying longer sequences of pulses to a two-level atom can produce a series of multi-temporal mode entangled states with properties intrinsically related to the Fibonacci sequence. Our results on photon-number entanglement can be} further exploited to generate new states of quantum light with applications in quantum technologies.
\end{abstract}
\maketitle

\noindent
Spontaneous emission is a phenomenon where an excited atom will spontaneously decay while emitting light into the vacuum of the electromagnetic field. Owing to its quantum coherent nature, spontaneous emission can preserve quantum properties such as entanglement and superposition. It has been used to prepare and measure atomic superposition states \cite{Monroe1131}, generate atom-atom \cite{ritter_elementary_2012,PhysRevLett.110.083603} and atom-photon entanglement \cite{blinov_observation_2004,wilk_single-atom_2007,de_greve_quantum-dot_2012}, create single photons \cite{brunel_triggered_1999,michler_quantum_2000}, and produce entangled photonic states by sequentially manipulating atomic systems \cite{neumann_multipartite_2008,cluster1D:Schwartz16,besse_realizing_2020}, as theoretically proposed in Refs. \cite{saavedra_controlled_2000,schon_sequential_2005,schon_sequential_2007}. It is key to developing quantum memories~\cite{PhysRevLett.79.769,specht_single-atom_2011} and quantum networks~\cite{PhysRevLett.78.3221,chou_functional_2007,yuan_experimental_2008,daiss_quantum-logic_2021}. 

\vspace{1mm}

As described in the seminal work of Weisskopf and Wigner \cite{Weisskopf:1930aa}, entanglement between light and matter naturally occurs during the spontaneous emission process of a two-level atom---an atom consisting of a ground $\ket{\mathrm{g}}$ and an excited $\ket{\mathrm{e}}$ state. Such entanglement lasts only until the spontaneous emission process brings the atom to the ground state, and as such, has not yet been considered as a direct resource for entangled light generation from a two-level system. Here, we show that the light-matter entanglement occurring during spontaneous emission can be controlled to produce photon-number entangled states distributed in the time domain. 

\vspace{1mm}

Consider a two-level atom with a spontaneous emission lifetime $T_1$. A short $\pi$-pulse excitation prepares the atom in the excited state $\ket{\mathrm{e}}$ at time $t=0$. In the absence of dephasing, at time $t > 0$, the light-matter system evolves into the entangled state $\alpha(t)\ket{\mathrm{e}}\ket{0}+\beta(t)\ket{\mathrm{g}}\ket{1}$, where $\alpha(t)=e^{-t/2T_1}$, $\beta(t)=\sqrt{1-\alpha(t)^2}$, and $\ket{n}$ is the state of emitted light containing $n$ photons. Thus, spontaneous emission from a two-level atom can be interpreted as a two-qubit gate that generates light-matter entanglement. For $t\gg T_1$, the atom is left in the state $\ket{\mathrm{g}}$, separable from the temporally-coherent single-photon state $\ket{1}$. 

\begin{figure}[t]
    \centering
    \includegraphics[width=0.48\textwidth,trim = 1mm 2mm 1mm 1mm, clip]{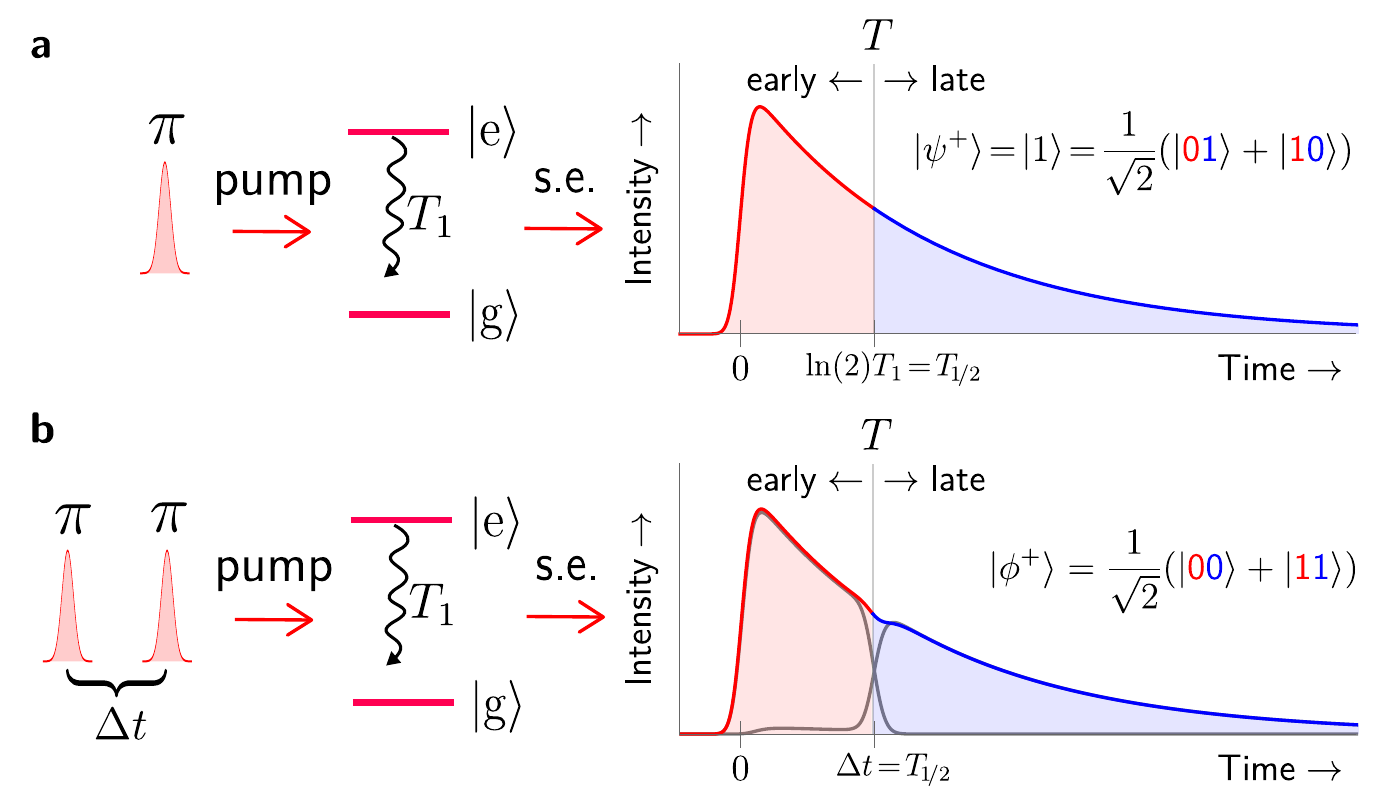}
    \caption{\textbf{Generation of photon-number Bell states.} \textbf{a}, A single photon is produced by spontaneous emission (s.e.) after a single $\pi$-pulse excitation of a two-level atom with a lifetime of $T_1$. Partitioning this photon into two orthogonal time bins (early and late) that are defined by setting the time bin threshold $T$ to the half-life $T_{\!1\!/2}=\ln(2)T_1$, reveals the photon-number $\ket{\psi^+}$ Bell state. \textbf{b}, Applying a subsequent $\pi$-pulse after time $\Delta t=T_{\!1\!/2}$ flips the state of the two-level atom while it is entangled with the photon field. By choosing $T$ to coincide with the second pulse, we find the Bell state $\ket{\phi^+}$.
    }
    \label{figure1}
\end{figure}

Along the temporal profile of the emitted single photon wavepacket, we can define adjacent early ($e$) and late ($l$) time bin modes separated by a chosen threshold time $T$, corresponding to the second-quantised temporal creation operators $\hat{t}^\dagger_e$ and $\hat{t}^\dagger_l$ in the pulse-mode formalism \cite{blow_continuum_1990,hussain_quantum_1992,ozdemir_pulse-mode_2002}. In this new time-bin basis, the pure single-photon state is now written as $\ket{1}=(\alpha(T) \hat{t}^\dagger_l+\beta(T)\hat{t}^\dagger_e)\ket{\mathbf{0}}=\alpha(T)\ket{0}_e\ket{1}_l+\beta(T)\ket{1}_e\ket{0}_l$. Note that, by choosing the time-bin threshold $T$ to be the half-life of the source $T_{\!1\!/2}=\ln(2)T_1$, the single-photon state is the photon-number Bell state $\ket{\psi^+}=(\ket{01}+\ket{10})/\sqrt{2}$ \cite{van2005single,specht_phase_2009}, where we have concatenated the time bins and dropped the subscripts for simplicity, see Fig. \ref{figure1}a. By extension, a single photon state could also be expressed as an $N$-mode $W$ state \cite{PhysRevA.62.062314} by conveniently defining $N$ time bin modes.

Now consider the application of a second $\pi$ pulse at time $\Delta t$ after the initial pulse, while the atom is still entangled with the field. This second pulse performs a single-qubit gate by coherently flipping the state of the atom so that, if a single photon was already emitted before $\Delta t$, it will emit a second photon after $\Delta t$, see Fig. \ref{figure1}b. Conversely, if no photon was emitted before the second pulse, then the atom is brought back to the ground state, preventing any emission from occurring. At time $\Delta t$ just after the second pulse, the total light-matter system is left in the entangled state $\alpha(\Delta t)\ket{\mathrm{g}}\ket{0}+\beta(\Delta t)\ket{\mathrm{e}}\ket{1}$. Hence, after emission has finished, the emitted photonic state becomes $(\alpha + \beta\hat{t}^\dagger_e\hat{t}^\dagger_l)\ket{\mathbf{0}}=\alpha\ket{0}_e\ket{0}_l+\beta\ket{1}_e\ket{1}_l$, as written in the photon number basis for time bins defined by a threshold time $T$ corresponding to the arrival of the second pulse $T=\Delta t$. Consequently, for $T=\Delta t =T_{\!1\!/2}$, the emitted photonic state is the photon-number Bell state $\ket{\phi^+}=(\ket{00}+\ket{11})/\sqrt{2}$. This simple approach can be scaled up to generate multi-mode entangled photonic states using multiple $\pi$ pulses, as discussed later on.

\vspace{3mm}
\noindent
\textbf{Results}
\vspace{1mm}

\noindent
We experimentally explore this scheme using a single semiconductor quantum dot acting as an artificial atom. The quantum dot is coupled to a micropillar cavity mode operating far into the bad-cavity regime~\cite{hilaire_deterministic_2020}, where emission into the cavity mode is irreversible. The device studied here consists of a negatively charged exciton addressed resonantly in a cross-polarised collection setup~\cite{RF:darkfield13} so that the optical transition is modelled as a resonantly driven two-level atom, which has a measured lifetime of $T_1=136\pm1$~ps. The laser excitation pulses are typically ten times shorter than the spontaneous emission lifetime of the transition and Rabi oscillations are observed as a function of the pulse power, attesting the coherent control of the device~\cite{PhysRevLett.87.133603}.

The single-photon nature of emission from the device is characterised by measuring a second-order correlation of $g^{(2)}(0)=0.063\pm0.002$ ($g^{(2)}$ for simplicity) after integrating over the pulsed emission following a single $\pi$-pulse excitation and normalising by the uncorrelated coincident counts at long delay times \mbox{$g^{(2)}_\tau>g^{(2)}$}. The coherent light-matter interaction during spontaneous emission is exemplified by the observation of Hong-Ou-Mandel (HOM) bunching \cite{HOM87,kiraz2004quantum} between successively emitted single-photons interfering at a beam splitter. We measure a correlation of $g^{(2)}_\text{HOM}=0.145\pm0.004$ at the output, attesting to the low probability for two photons to exit the beam splitter separately. These measurements together provide a mean wavepacket overlap of $M=1-2g^{(2)}_\text{HOM}+g^{(2)}=0.77\pm0.01$ and an estimated single-photon indistinguishability $M_\text{s}=M/(1-g^{(2)})=0.82\pm0.02$ at the source \cite{ollivier_hong-ou-mandel_2021}. 

The indistinguishability of a single photon wavepacket characterises how coherent it is in time. The $\ket{\psi^+}$ Bell state is also strongly linked to this same temporal coherence, being a superposition of states $\ket{01}$ and $\ket{10}$ of a photon arriving in two different time bins. In the supplementary, we theoretically show and experimentally verify that the Bell-state fidelity of a single photon with respect to $\ket{\psi^+}$ is well-approximated by $\mathcal{F}_{\psi^+}\simeq p_1\sqrt{M_\mathrm{s}}=0.88\pm0.02$ when choosing $T=T_{\!1\!/2}=94$~ps, where $p_1$ is the probability of emitting a single photon. Here, we proceed to experimentally explore the proposed scheme to generate the $\ket{\phi^+}$ Bell state by applying a second $\pi$ pulse separated from the first by the half-life $\Delta t\simeq T_{\!1\!/2}$.

The ideal $\ket{\phi^+}$ state is composed of two photons with a probability of $p_2=1/2$ and the vacuum otherwise ($p_0=1/2$). This renders an expected intensity correlation of $g^{(2)}=1$ and an average photon number of $\mu=1$ at the source. We confirm this prediction by measuring $g^{(2)}=0.99\pm0.02$ and $\mu/\mu_\pi=1.02\pm0.01$ with respect to the average photon number $\mu_\pi$ produced by a single pulse, which is expected to be near unity at the source. We also verify that producing three or more photons is rare by measuring a small third-order correlation $g^{(3)}=0.165\pm0.007$, corresponding to a three-photon emission probability of about $3\%$. A detailed discussion of photon number probabilities and losses is given in the supplementary. Thus, the photon statistics already suggest a state of the form $\ket{0}+\ket{2}$. It remains now to demonstrate a separation of the two photons into an early and a late time bin $\ket{11}$ and the presence of a coherence with the vacuum part of the state $\ket{00}$.

\begin{figure}[t]
	\centering
	\includegraphics[width=0.485\textwidth,trim=1mm 2mm 1mm 1mm, clip]{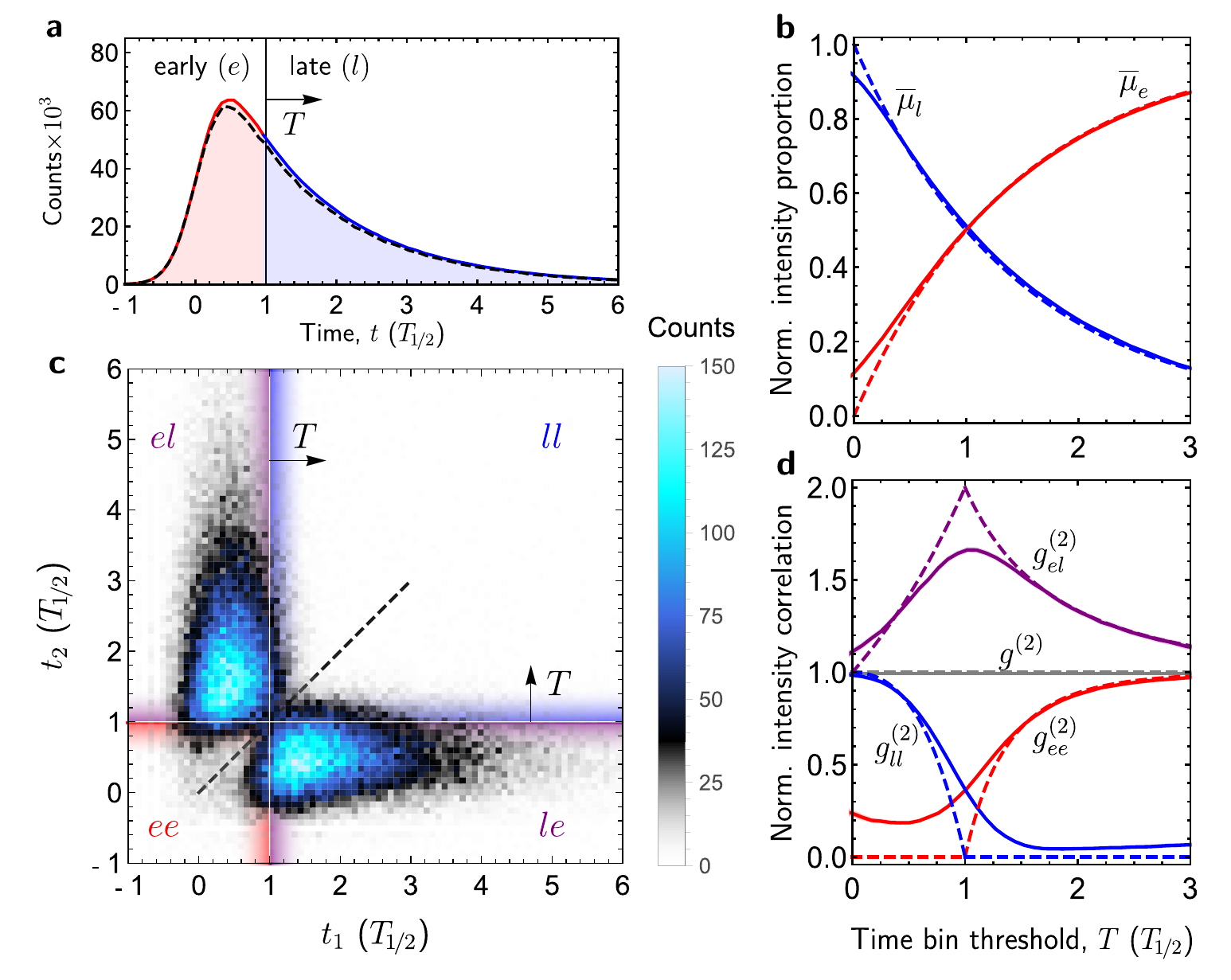}\vspace{0mm}
	\caption{\textbf{Characterisation of intensity.} 
\textbf{a}, The temporal profile measured after applying two $\pi$ pulses separated in time by the half-life $\Delta t{\simeq}T_{\!1\!/2}$. It is divide into early ($e$, red) and late ($l$, blue) bins defined by the threshold time $T$. The black dashed line shows the single-photon profile obtained for the same measurement duration after applying a single $\pi$ pulse. \textbf{b}, The normalised proportion of counts detected in the early $\overline{\mu}_e$ and late $\overline{\mu}_l$ time bins as $T$ is swept over the temporal profile. \textbf{c}, The time-resolved intensity correlation map $G^{(2)}(t_1,t_2)$ divided by $T$ into four quadrants corresponding to each pair of time bins where a coincidence detection occurs. \textbf{d}, The intensity correlation for each quadrant $g^{(2)}_{ab}$ normalised by the square average photon number $\mu_a\mu_b$ detected in bins $a,b\in\{e,l\}$, computed from panel \textbf{c} as $T$ is swept across the wavepacket. We obtain $g^{(2)}_{el}=g^{(2)}_{le}$ by averaging the counts in the off-diagonal quadrants. The dashed curves in panels \textbf{b} and \textbf{d} show the values expected for an ideal $\ket{\phi^+}$ Bell state. The solid curves show the measured values where the standard uncertainty is smaller than the thickness of the line.
	}
\label{figure2}
\end{figure}

To access temporal properties after the application of two pulses, we first measure the temporal profile and find that it matches well to the profile produced after excitation by a single $\pi$ pulse (see Fig.~\ref{figure2}a), as foreshadowed by Fig.~\ref{figure1}. By sweeping the time-bin threshold $T$ across the wavepacket, we find that the proportion of counts $\overline{\mu}_a=\mu_a/\mu$ detected in each time bin $a\in\{e,l\}$ cross at the half-life condition (see Fig.~\ref{figure2}b). This matches the trend given by the ideal $\ket{\phi^+}$ Bell state.

We study the two photons composing the total temporal profile by performing time-resolved intensity correlation measurements (see \emph{Methods}). This produces a two-time coincidence map $G^{(2)}(t_1,t_2)$ that can be divided into four time bin quadrants defined by a chosen $T$, designated $ee$, $el$, $le$, and $ll$ as shown in Fig.~\ref{figure2}c. The direct inspection of this map reveals that coincident counts between different time bins ($el$, $le$) predominantly occur when $T=\Delta t\simeq T_{\!1\!/2}$, indicating that the two photons are indeed temporally separated.

To quantify this observation, we analyse each quadrant of the $G^{(2)}$ map individually. The counts in each quadrant are summed and normalised by the product of average photon numbers $\mu_a\mu_b$ obtained within each pair of bins $a,b\in\{e,l\}$. This gives the normalised correlation $g^{(2)}_{ab}$ for the pair of time bins where a coincidence count was detected. The total $g^{(2)}$ is then seen as an average of each $g^{(2)}_{ab}$ weighted by the proportion $\overline{\mu}_a\overline{\mu}_b$.

The time bin analysis of $g^{(2)}$, presented in Fig.~\ref{figure2}d, reveals that anti-bunching occurs for detection within the same bins ($g^{(2)}_{ee}, g^{(2)}_{ll}< 1$), whereas bunching occurs between different bins ($g^{(2)}_{el}=g^{(2)}_{le}> 1$). The bunching is maximum when the time-bin threshold is chosen at the half-life. However, the amount of bunching is less than would be expected from an ideal state produced by infinitesimally short pulses and measured with perfect time resolution (dashed curves). This is primarily due to the detection time jitter (see \emph{Methods}), which occasionally detects photons in quadrants $ee$ and $ll$ that would otherwise reside in $el$ or $le$ and hence decreases $g^{(2)}_{el}$ while increasing $g^{(2)}_{ee}$ and $g^{(2)}_{ll}$. From this intensity correlation analysis, we find that $81.5\pm0.4\%$ of two-photon measurements occur in different time bins, evidencing a primary $\ket{11}$ component. We now probe the expected coherent properties of the photonic state using phase correlation measurements.

\begin{figure*}[t]
    \centering
    \includegraphics[width=0.98\textwidth,trim=1mm 1mm 1mm 1mm, clip]{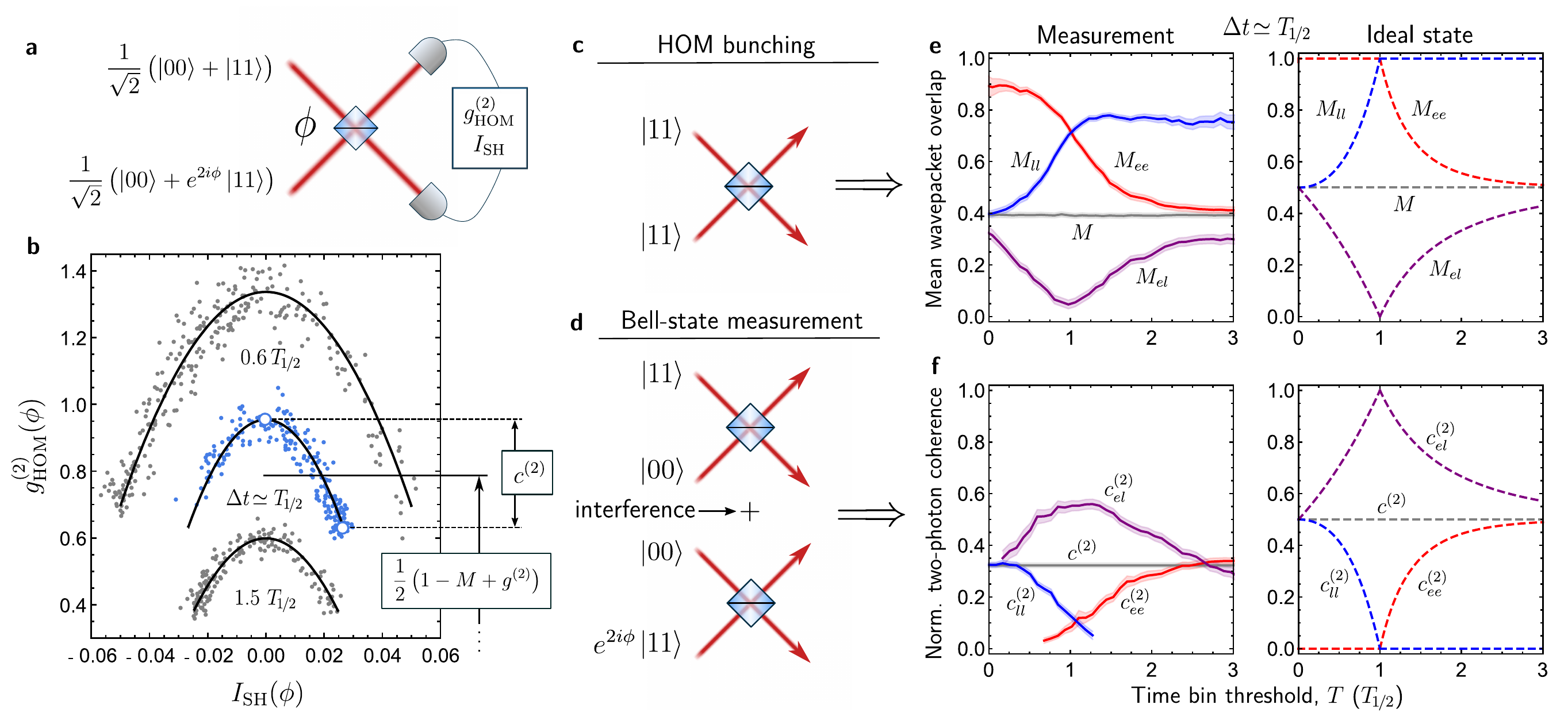}
    \caption{
    \textbf{Characterisation of coherence.} 
    \textbf{a}, Two ideal Bell states interfering at a beamsplitter, after which the normalised intensity correlation $g^{(2)}_\text{HOM}$ and self-homodyne signal $I_\text{SH}$ are measured.
    \textbf{b}, The fitted quadratic relationship (black curve) between the coincidence counts ($g^{(2)}_\text{HOM}(\phi)$) and the detection rate difference ($I_\text{SH}(\phi)$) for three different pulse separations $\Delta t$ (see labels, data for $1.5T_{\!1\!/2}$ are shifted down by 0.2 for clarity). Each data point in this panel is computed by integrating over the entire time-resolved correlation map $G^{(2)}_\text{HOM}(t_1,t_2,\phi)$ for a given measured $I_\text{SH}(\phi)$.
    \textbf{c}, Sketch of the HOM bunching case when interfering two $\ket{11}$ states. \textbf{d}, Sketch of the path-entangled Bell state interference.
    \textbf{e}, The mean wavepacket overlaps $M_{ab}$ and, \textbf{f}, normalised second-order coherence magnitudes ${c}^{(2)}_{ab}$ for time bins $a,b\in\{e,l\}$. These quantities are extracted as in panel \textbf{b} but using instead $g^{(2)}_{\text{HOM},ab}(\phi)$ obtained by integrating and normalising $G^{(2)}_\text{HOM}(t_1,t_2,\phi)$ for each quadrant, as was done for $G^{(2)}(t_1,t_2)$ to obtain $g^{(2)}_{ab}$ in Fig.~\ref{figure2}. The shaded regions show the standard uncertainty obtained from fitting the scattered data. For clarity, we show separate panels for the values expected of an ideal $\ket{\phi^+}$ Bell state (dashed curves).
    }
    \label{figure3}
\end{figure*}

The intensity correlation $g^{(2)}_\text{HOM}$ at the output of a path-unbalanced Mach-Zehnder interferometer, commonly used to measure HOM bunching, can oscillate as the interferometer phase $\phi$ evolves. This occurs when the input contains coherence between any states differing by two photons (second-order coherence). We recently used this technique to measure the amount of coherence generated between the vacuum and two photons when exciting a two-level atom with a single 2$\pi$ pulse \cite{loredo_generation_2019}. We use this same concept to characterise the number coherence generated after applying sequential $\pi$ pulses by interfering two of the generated photonic states (see Fig.~\ref{figure3}a).

The correlation $g^{(2)}_\text{HOM}$ depends on both $g^{(2)}$ and the mean wavepacket overlap $M$ \cite{ollivier_hong-ou-mandel_2021}. By considering photon-number coherence, a phase-dependent term arises \cite{scweinthesis}:
\begin{equation}
    2g^{(2)}_\text{HOM}(\phi)=1-M+g^{(2)}-{c}^{(2)}\cos(2\phi),
\end{equation}
where ${c}^{(2)}$ is an intensity-normalised value quantifying the second-order coherence, as described in the supplementary. In our setup, $\phi$ freely evolves on a slow timescale (see \emph{Methods}). To accurately extract ${c}^{(2)}$, we simultaneously monitor the self-homodyne signal ${I_\text{SH}=(\mu^+-\mu^-)/\mu\propto\cos(\phi)}$, which is the normalised difference in average photon number $\mu^\pm$ detected at each output. Since $g^{(2)}_\text{HOM}$ depends on the phase through $-\cos(2\phi)$ and $I_\text{SH}$ through $\cos(\phi)$, we expect a quadratic phase-correlated parametric relationship $g^{(2)}_\text{HOM}\sim-I_\text{SH}^2$ with an amplitude of ${c}^{(2)}$.

Although an ideal $\ket{\phi^+}$ state should give $I_\text{SH}=0$, as it does not have first-order coherence \cite{loredo_generation_2019}, the finite temporal width of pulses applied to the atom inevitably cause a small signal $|I_\text{SH}|\ll1$. We believe this signal is produced by the atom directly when a photon is occasionally emitted during the excitation pulse, which allows for the remainder of the pulse to prepare the atom in a superposition state. However, it could also arise from over/under-estimating the $\pi$ pulse conditions or from imperfect polarisation filtering of the excitation pulses. By monitoring this remnant self-homodyne signal, we observe the expected quadratic signature and use it to measure ${c}^{(2)}$ for three different pulse separations $\Delta t$ (see Fig.~\ref{figure3}b). The amplitude of oscillation increases with decreasing $\Delta t$ due to normalising by intensity, hence illustrates a convergence toward the vacuum. A full analysis and discussion of measurements when varying $\Delta t$ is available in the supplementary. The total time-integrated value ${c}^{(2)}$ indicates significant second-order coherence, but it does not distinguish states of the form $\ket{0}+\ket{2}$ from $\ket{00}+\ket{11}$. For this, we resolve the measurement in time.

Consider the interference of two ideal Bell states (recall Fig.~\ref{figure3}a). This gives rise to four cases. First, the vacuum inputs $\ket{00}$ will give a trivial output. Second, two $\ket{11}$ states will cause HOM bunching (see Fig.~\ref{figure3}c). As opposed to the ideal single-photon case where $M=1$, an ideal $\ket{\phi^+}$ state should give $M=1/2$. This is because both early and late photons bunch with their pair in the same time bin ($M_{ee}=M_{ll}=1$), but each pair can still exit the beam splitter independently ($M_{el}=M_{le}=0$). Third, the cases combining $\ket{00}$ and $\ket{11}$ can occur in two ways (see Fig.~\ref{figure3}d). If these latter two cases produce indistinguishable outputs, then a quantum interference occurs due to the erasure of the information about which path the two photons took through the interferometer. This two-photon interference evidences the presence of a path-entangled Bell state between the upper $U$ and lower $L$ paths of the interferometer: $(\ket{U}_e\ket{U}_l+e^{2i\phi}\ket{L}_e\ket{L}_l)/\sqrt{2}$, and it causes an oscillation of coincident counts depending on $\phi$ that contributes to the ${c}^{(2)}$ term of $g^{(2)}_\text{HOM}$. Thus, monitoring the oscillation of coincidence counts constitutes a Bell-state measurement of this path-entangled state produced by the $\ket{\phi^+}$ input. However, to distinguish $\ket{\phi^+}$ from arbitrary states of the form $\ket{0}+\ket{2}$, we must measure the component of ${c}_{el(le)}^{(2)}$ arising from coincidences between photons arriving in different time bins and show that it exceeds any contribution from ${c}^{(2)}_{ee(ll)}$.

To measure $M_{ab}$ and ${c}^{(2)}_{ab}$, we use the same approach used to obtain $g^{(2)}_{ab}$, by analysing each quadrant of the time-resolved correlation map $G^{(2)}_\text{HOM}(t_1,t_2,\phi)$ for a given threshold $T$. We subdivide each map $G^{(2)}_\text{HOM}(t_1,t_2,\phi)$ corresponding to each $I_\text{SH}(\phi)$. This produces four quadratic signatures similar to those presented in Fig.~\ref{figure3}b, one corresponding to each quadrant. We then fit these four sets of data to extract the quantities ${c}^{(2)}_{ab}$ and $M_{ab}$.

From the time bin analysis of phase correlations, we see that the mean wavepacket overlaps of photons in the same bins $M_{ee}$ and $M_{ll}$ both remain relatively high and intersect at the half-life, while the overlap between bins $M_{el}$ dips nearly to zero (see Fig.~\ref{figure3}e). This indicates that the photons composing $\ket{11}$ are mostly individually indistinguishable, yet almost fully distinguishable from each other. Interestingly, $M_{ee}$ exceeds the mean wavepacket overlap measured after a single $\pi$ pulse ($M\simeq0.77$) when $T<T_{\!1\!/2}$. We attribute this to the sharp temporal truncation of photons in the early bin, which causes a spectral broadening that partially overcomes dephasing. This truncation does not modify the temporal shape of photons in the late bin---they remain exponentially-decaying profiles. Hence, $M_{ll}$ converges to the single-photon case when $T>T_{\!1\!/2}$. The observed crossing and dip follows that predicted by the ideal state and verifies the scenario described by Fig.~\ref{figure3}c when $T=T_{\!1\!/2}$.

We find that the trend for ${c}^{(2)}_{ab}$ mimics that of $g^{(2)}_{ab}$, as predicted by the ideal state, with ${c}^{(2)}_{el}$ peaking when ${c}^{(2)}_{ee}$ and ${c}^{(2)}_{ll}$ intersect at the half-life (see Fig.~\ref{figure3}f). However, the magnitudes are further suppressed relative to the ideal case because the coherence is susceptible to dephasing in addition to errors caused by imperfect pulses and detection jitter. That said, we find that ${c}^{(2)}_{el}$ is much greater than ${c}^{(2)}_{ee}$ and ${c}^{(2)}_{ll}$ at the half-life, indicating that the majority of the oscillation observed in $g^{(2)}_\text{HOM}$ arises from a coherence between the vacuum $\ket{00}$ and two photons arriving in orthogonal time bins $\ket{11}$.

The three intensity-normalised quantities ${c}^{(2)}_{el}$, $M_{ee}$, and $M_{ll}$ can be used to estimate the magnitude of some density matrix elements of the photonic state at the source, before losses from collection. From this, we estimate that the emitted state has an entanglement concurrence of $\mathcal{C}=0.70\pm0.05$. Note that a positive value $\mathcal{C}>0$ unambiguously indicates the presence of quantum entanglement \cite{wootters2001entanglement}. We also estimate a fidelity of $\mathcal{F}_{\phi^+}=0.79\pm0.03$ with respect to the $\ket{\phi^+}$ Bell state.

\vspace{4mm}
\noindent
\textbf{Discussion}
\vspace{1mm}

\noindent
Our fidelity and concurrence estimates are limited by the detector jitter time. The measurements of total second-order coherence ${c}^{(2)}$ and mean wavepacket overlap $M$ suggest that a fidelity up to $0.86$ is possible with this device using detectors with a time jitter well below \cite{korzh_demonstration_2020} the pulse timescale ($t_\mathrm{p}=20$~ps) used in our experiments, which would reduce the proportion of two-photon events occurring within the same time bins ($ee$ or $ll$) down to $3t_\mathrm{p}/8T_1\simeq 5\%$ (see supplementary). Using shorter pulses ($t_\mathrm{p}\ll T_1$), when combined with low time-jitter detectors, would bring the fidelity up to at most $\sqrt{M_\mathrm{s}}\simeq0.91$, which is limited by the dephasing of this device. Note that $M_\mathrm{s}\geq0.975$ has been achieved with quantum dot devices \cite{SPS:Somaschi16,tomm2021bright}, which could provide a fidelity up to 0.987.

In our experiments, we do not perform a full quantum state tomography on the photonic state to retrieve its density matrix because it is difficult to spatially separate and independently analyse the time bin modes. We instead characterise the photon-number entanglement of the state via intensity and phase correlation measurements, which under some reasonable assumptions allows for a partial reconstruction of the density matrix and for estimates of fidelity and concurrence (see supplementary). One approach to separate time bins would be to use an ultrafast optical switch, which would then allow for single-qubit gates, quantum teleportation, and Bell tests. For our system, the short lifetime dictates an optical switching time on the picosecond timescale, which is achievable using lithium niobate integrated photonic circuits \cite{wang_integrated_2018}. However, our approach for generating photonic entanglement can be applied to any coherently controlled source of indistinguishable photons modelled by a two-level system.

Our entangling protocol also has a simple extension to multimode entanglement by applying a longer sequence of $\pi$ pulses. As detailed in the supplementary, the photonic state $\ket{\psi_N}$ produced by $N$ pulses has a recursive nature that becomes transparent when labelling the time bins in reverse chronological order. In this case, by applying the matrix product state formalism \cite{schon_sequential_2005,schon_sequential_2007}, we find that the final state can be determined from the Fibonacci-like relation
\begin{equation}
\label{fibonacciEq}
\ket{\psi_N} = \alpha_{N}\ket{\psi_{N-2}}+\beta_{N}\hat{t}^{\dagger}_N\ket{\psi_{N-1}},
\end{equation}
where $\alpha_m=e^{-\Delta t_m/2T_1}$, $\beta_m=\sqrt{1-\alpha^2_m}$, $\hat{t}^\dagger_m\ket{0}_m=\ket{1}_m$, and where $m$ labels the $m$th time bin from the end of the sequence. If $\Delta t_1\gg T_1$ so that the atom relaxes to the ground state at the end of the sequence, then $N=1$ pulse produces a single photon $\ket{\psi_1}=\ket{1}_1$ and $N=2$ pulses produces the entangled state: $\ket{\psi_2}=\alpha_2\ket{0}_2\ket{0}_1+\beta_2\ket{1}_2\ket{1}_1$. By choosing the pulse separations $\Delta t_2=T_1\ln(2)$ and $\Delta t_3=T_1\ln(3)$, we obtain the maximally entangled $W$-class state produced by $N=3$ pulses: $\ket{\psi_3}=(\ket{001}+\ket{100}+\ket{111})/\sqrt{3}$. In general, the entangled states produced by this sequence belong to the class of matrix product states with 2-dimensional bonds \cite{schon_sequential_2005,schon_sequential_2007}, and they are not equivalent to $N$-qubit $W$ states for $N\geq 4$. Further studies are needed to identify the type and amount of entanglement provided by multi-pulse sequences applied to two-level atoms.

\vspace{4mm}
\noindent
\textbf{Conclusions}
\vspace{1mm}

\noindent
We have shown that the light-matter entanglement occurring during spontaneous emission from a two-level atom is a fundamental resource for generating entangled light. By probing the temporal domain of pulsed light emitted by an artificial atom after a double $\pi$-pulse excitation, our measurements demonstrate the generation of a photon-number Bell state. By adding more consecutive $\pi$-pulses, we herald that this protocol can produce multipartite temporal entanglement, and it is a step closer to the generation of high-order Fock states and cat states, which require dynamic control of the light-matter coupling strength \cite{PhysRevLett.76.1055,PhysRevResearch.2.033489,PhysRevResearch.3.023088}. Such a new class of photonic states could serve as building blocks for distributing entanglement, quantum state teleportation, and may allow new ways to implement quantum random walks, quantum sensing, and photonic networks \cite{briegel_persistent_2001}. We also believe that the sequential coherent driving of multi-level atomic systems during spontaneous emission offers promising perspectives for generating high-dimensional entanglement \cite{erhard_advances_2020}; for example, using the biexciton-exciton cascade in semiconductor quantum dots or in combination with spin-photon entanglement protocols.

\vspace{4mm}
\noindent
\textbf{Acknowledgements}
\vspace{1mm}

\noindent
P.S. acknowledges support from the ERC PoC PhoW, the IAD - ANR support ASTRID program Projet Grant. No. ANR-18-ASTR-0024 LIGHT, the QuantERA ERA-NET Cofund in Quantum Technologies, project HIPHOP, the FET OPEN QLUSTER, the French RENATECH network, a public grant overseen by the French National Research Agency (ANR) as part of the ``Investissements d’Avenir” programme (Labex NanoSaclay, Grant No. ANR-10-LABX-0035). J.C.L. acknowledges the Austrian Federal Ministry for Digital and
Economic Affairs, the National Foundation for Research, Technology and Development, and the Christian Doppler Research Association. A.A., M.M. and S.C.W acknowledge support from the Foundational Questions Institute Fund Grant No. FQXi-IAF19-01 (A.A and S.C.W) and FQXi-IAF19-05 (A.A. and M.M.), as well as the European Union’s Horizon 2020 research and innovation programme under the Marie Skłodowska-Curie grant agreement No. 861097 (A.A., M.M., and S.C.W.). A.A. and M.M also acknowledge support from the ANR Research Collaborative Project ``Qu-DICE" Grant No. ANR-PRC-CES47 (A.A and M.M.), the Templeton World Charity Foundation Inc Grant No. TWCF0338 (A.A. and M.M.), and the John Templeton Foundation Grant No. 61835 (A.A.). C.S. acknowledges support from the Natural Sciences and Engineering Research Council of Canada (NSERC) Discovery Grant and Strategic Project Grant programs, and also the National Research Council's High-Throughput Secure Networks program. S.C.W. also acknowledges support from the NSERC Canadian Graduate Scholarships (Grant No. 668347 and 677972) and the SPIE Education Scholarship program. S.C.W. and C.A-S. acknowledge A. Gonz\'alez-Tudela, C. S\'anchez-Mu\~noz, T. Huber and N. Sinclair for fruitful discussions. C.A-S. thanks K. Bencheikh and F. Raineri for providing technical assistance in the experimental setup.

\vspace{3mm}
\noindent
\textbf{Author contributions}
\vspace{1mm}

\noindent
The experiments were conducted by J.C.L. and C.A-S. Data analysis was carried out by S.C.W. and C.A-S. with help from J.C.L. and P.H. The theoretical modelling was done by S.C.W., M.M., C.S. and A.A, with help from J.C.L and C.A-S. The cavity devices were fabricated by A.H. and N.S. from samples grown by A.L., based on a design of L.L.; the etching was done by I.S. The manuscript was written by S.C.W. and C.A-S. with assistance from C.S. and P.S. and input from all authors. The project was supervised by C.A-S. with the collaboration of C.S. and P.S.

\vspace{3mm}
\noindent
\textbf{Competing interests}
\vspace{1mm}

\noindent
N.S. is co-founder, and P.S. is scientific advisor and co-founder, of the single-photon-source company Quandela. The remaining authors declare no competing interests.

\vspace{3mm}
\noindent
\textbf{Data availability}
\vspace{1mm}

\noindent
The experimental data that support the findings of this study are available in figshare at \href{https://doi.org/10.6084/m9.figshare.16838248}{doi.org/10.6084/m9.figshare.16838248}.

\vspace{3mm}
\noindent
\textbf{Methods}
\vspace{1mm}

\noindent
More details about the QD-micropillar source used in our experiments can be found in Ref. \cite{Ollivier:2020aa}, source number 3. This source was chosen over others for its very high emission efficiency, which allows for fast collection of time-resolved maps and third-order intensity correlations. The experiments were performed in a standard resonant cross-polarisation setup \cite{RF:darkfield13,loredo_generation_2019,Ollivier:2020aa}. We prepare the two $\pi$-pulse sequence in a compact Michelson interferometer. This provides passive phase-stabilisation for the delayed output pulses and independent intensity tuning. One of the mirrors is mounted on a nanometric translation stage allowing for delay tuning up to $175$~ps. The laser pulses have a temporal FWHM of ${\sim}20$ ps. The coincidence maps are retrieved via time-tagging of the photon events with respect to the laser clock \cite{flagg_dynamics_2012}. The phase correlation measurements are implemented in a path unbalanced Mach-Zehnder interferometer, with a delay of 12.3 ns in one of the arms, matching the laser repetition rate\cite{loredo_generation_2019}. The phase $\phi$ of the interferometer evolves slowly, performing a $\pi$ shift on the ${\sim}5$ s timescale. The relatively fast 100 ms time to acquire a single time-resolved correlation map $G^{(2)}_\text{HOM}(t_1,t_2,\phi)$ allows us to consider the phase $\phi$ constant for each map. 

\vspace{1mm}

The device and setup losses are detailed in Refs. \cite{hilaire_deterministic_2020,Ollivier:2020aa}. The probability of having a single photon per pulsed excitation in the collection single-mode fiber is ${\sim}10\%$. All measurements are performed using superconducting nanowire single-photon detectors with ${\sim}70\%$ quantum efficiency and ${\sim}50$ ps FWHM Gaussian jitter time. The $g^{(3)}$ and $g^{(2)}_\text{HOM}$ measurements are intensity-normalised quantities that are insensitive to photon losses, thus allowing to characterise the photonic state at the source level, before any photon collection, transmission, and detection losses. Under a single $\pi$-pulse excitation, the count rate per detector in the intensity correlation measurements ($g^{(3)}$ Hanbury-Brown-Twiss setup, with three detectors) is $1.256\pm0.007$ MHz/detector, and the count rate per detector for the $g^{(2)}_\text{HOM}$ coherence measurements (output of the Mach-Zehnder interferometer) is $0.899\pm0.005$ MHz/detector. The photon time-tags are processed in a HydraHarp 400 autocorrelator, with a temporal discretisation of 8 ps.

\let\oldaddcontentsline\addcontentsline
\renewcommand{\addcontentsline}[3]{}

 \bibliographystyle{naturemag}

\let\addcontentsline\oldaddcontentsline

\clearpage
\onecolumngrid

\begin{center}
  \textbf{\large Supplementary Material: Photon-number entanglement generated by sequential excitation of a two-level atom}\\[.5cm]
  Stephen C. Wein,$^{1, 2,*}$ Juan C. Loredo,$^{3,4,5}$ Maria Maffei,$^{2}$ Paul Hilaire,$^{3,6}$ Abdelmounaim Harouri,$^{3}$ \\Niccolo Somaschi,$^{7}$ Aristide Lema\^itre,$^{3}$ Isabelle Sagnes,$^{3}$ Lo\"ic Lanco,$^{3,8}$ Olivier Krebs,$^{3}$ \\Alexia Auff\`eves,$^{2}$ Christoph Simon,$^{1}$ Pascale Senellart,$^{3}$ and Carlos Ant\'on-Solanas$^{3, 9,\dagger}$
  \\[.1cm]
  {\itshape 
 
$^1$Institute for Quantum Science and Technology and Department of Physics and Astronomy, \\University of Calgary, Calgary, Canada T2N 1N4

$^2$Universit\'{e} Grenoble Alpes, CNRS, Grenoble INP, Institut N\'eel, 38000 Grenoble, France

$^3$Centre  for  Nanosciences  and  Nanotechnology,  CNRS,  Universit\'e  Paris-Saclay,\\UMR  9001,  10  Boulevard  Thomas  Gobert,  91120  Palaiseau,  France

$^4$University of Vienna, Faculty of Physics, Vienna Center for \\Quantum Science and Technology (VCQ), Vienna, Austria

$^5$Christian Doppler Laboratory for Photonic Quantum Computer, Faculty of Physics, \\University of Vienna, 1090 Vienna, Austria

$^6$Department of Physics, Virginia Tech, Blacksburg, Virginia 24061, USA

$^7$Quandela  SAS,  10  Boulevard  Thomas  Gobert,  91120  Palaiseau,  France

$^8$Universit\'e de Paris, Centre for Nanoscience and Nanotechnology (C2N), F-91120 Palaiseau, France

$^9$Institute  of  Physics,  Carl  von  Ossietzky  University,  26129  Oldenburg,  Germany \\ }
\end{center}

\setcounter{equation}{0}
\setcounter{figure}{0}
\setcounter{table}{0}
\setcounter{page}{1}
\makeatletter
\renewcommand{\theequation}{S\arabic{equation}}
\renewcommand{\thefigure}{S\arabic{figure}}
    In this document, we describe the details of the theoretical and experimental analysis of photon-number entangled states generated from a two-level atom. In section \ref{supsection:theorymodel}, we derive the model of photon-number entanglement from the light-matter interaction of a two-level atom with the photonic field. We then describe, in section \ref{supsection:theorymeasurement}, the theoretical details for the measurements we perform, assuming generally imperfect photonic states. In section \ref{supsection:experimental}, we present the extended experimental analysis of the generated photonic states: the single photon Bell state $\ket{\psi^+}$, as well as 9 different pulse separations for the double $\pi$ pulse sequence, which we used to produce the $\ket{\phi^+}$ state at the half-life condition presented in the main text.

\tableofcontents

\clearpage
\section{Theoretical model of the joint emitter-field dynamics}
\label{supsection:theorymodel}

We consider a two-level atom system coupled to a multimode electric field in a one-dimensional waveguide. The free Hamiltonian of the atom is $H_\text{s}= \hbar \omega_0 (\hat{\sigma}_z+\hat{1})/2$, with $\hat{\sigma}_z= \ket{\mathrm{e}}\bra{\mathrm{e}}-\ket{\mathrm{g}}\bra{\mathrm{g}}$, where $\ket{\mathrm{e}}$ ($\ket{\mathrm{g}}$) is the excited (ground) state. The free Hamiltonian of the field is $H_\text{F}=\sum_k \hbar\omega_k \au_k \ad_k$, where $\au_{k}$ ($\ad_k$) creates (annihilates) a photon of frequency $\omega_k$; we consider that the light propagates only in one direction and with constant velocity $v$, so that $\omega_k= v k$. We also assume that the coupling strength between the emitter and field is uniform in frequency. Then, the interaction potential reads $V= i \hbar g \sum_k(\hat{\sigma}_{-} \au_k -\hat{\sigma}_{+} \ad_k)$, with $\hat{\sigma}_{-}=\ket{\mathrm{g}}\bra{\mathrm{e}}$ and $\hat{\sigma}_{+}=\ket{\mathrm{e}}\bra{\mathrm{g}}$. We study the dynamics in the interaction picture: the operators evolve with $H_{0}=H_\text{s} + H_\text{F}$, and the states with $V(t)= i\hbar g \sum_k(e^{-i(\omega_0  -\omega_k) t}\hat{\sigma}_{-} \au_k - e^{i(\omega_0  -\omega_k) t}\hat{\sigma}_{+} \ad_k)$.

We fix a time interval $\delta t$, such that $t_n= n \delta t$, then we define a new operator acting on the field's temporal modes: $\bd_n=\sqrt{\delta t/\rho_{\omega}}\sum_k e^{-i \omega_k t_n }\ad_k$, where $\rho_{\omega}$ is the uniform density of waveguide modes and we have that $[\bd_{n},\bu_{n^{\prime}}]=\delta_{n,n^{\prime}}$. For $\delta t$ small enough, $\int_{t_n}^{t_{n+1}} dt' V(t')\approx \delta t V(t_n)$, with $V(t_n)= i\hbar \sqrt{\gamma/\delta t}(e^{-i\omega_0 t_n}\hat{\sigma}_{-} \bu_n - e^{i\omega_0 t_n}\hat{\sigma}_{+} \bd_{n} )$, where $\gamma= g^2 \rho_{\omega}$. Then, we can imagine that, between time $t_n$ and $t_{n+1}$, the emitter interacts only with the $n$-th discretised temporal mode of the field, and the global system's state evolves under the unitary transformation $U_{n}=e^{\frac{-i}{\hbar}\delta t V(t_n)}$. Noticing that, for every $n$ and $n^{\prime}$ with $n\neq n^{\prime}$, $[U_{n},U_{n'}]=0$, we can factor the unitary evolution between time $0$ and $t_N$. Furthermore, taking $\delta t \ll \gamma^{-1}$, we can approximate $U_{n}$ up to its first order in $\gamma\delta t$, ${\cal U}_n $. For example, if the initial state is $\ket{\mathrm{e},\textbf{0}}$, where $\ket{\textbf{0}}\equiv\bigotimes_n\ket{0}_n$ is the field vacuum, at time $t_N$ the state is $\ket{\Psi(t_N)}=\mathcal{U}_{N-1}\mathcal{U}_{N-2}...\mathcal{U}_{0}\ket{\mathrm{e},\textbf{0}}$. Expanding this expression and taking the continuous limit, $\sum_n \delta t \rightarrow \int dt$, $\bu_n\rightarrow \sqrt{\delta t}\au(t)$ we get
\begin{equation}
\label{Eq:amplitudedamping}
\ket{\Psi(t)}=e^{-\gamma t/2}\ket{\mathrm{e},\textbf{0}}+\sqrt{\gamma}\int_{0}^{t} dt^{\prime} e^{-  \gamma t^{\prime}/2}e^{- i\omega_0 t^{\prime}}\au(t^{\prime})\ket{\mathrm{g},\textbf{0}}
\equiv \sqrt{e^{-\gamma t}} \ket{\mathrm{e},\textbf{0}}+\sqrt{1-e^{-\gamma t}}\ket{\mathrm{g},\textbf{1}},
\end{equation}
where $\ket{\textbf{1}}$ is the normalised state of the field containing one photon.

We can now find the total wavefunction when the atom is driven by a resonant coherent pulse. We consider the case of a square pulse of amplitude $\alpha_0$ and duration $t_\mathrm{p}$. For each $t_n \in [0,t_\mathrm{p}]$, the single step unitary operator is now $\tilde{U}_n=e^{\frac{-i }{\hbar}\delta t\tilde{V}(t_n)}$. Where $\tilde{V}(t_n)=V(t_n)+\hbar(\Omega/2) \hat{\sigma}_y $, with Rabi frequency $\Omega =2\alpha_0\sqrt{\gamma/\rho_{\omega}} $.  We take again first order in $\gamma \delta t$, $\tilde{{\cal U}}_{n}$. If the initial state of the atom is $\ket{j}$, with $j\in\{\mathrm{g},\mathrm{e}\}$, at time $t_N\leq t_\mathrm{p}$, the global system's state is given by $\ket{\tilde{\Psi}(t_N)}=\tilde{{\cal U}}_{N-1}\tilde{{\cal U}}_{N-2}...\tilde{{\cal U}}_{0}\ket{j,\textbf{0}}$. Expanding this expression and taking the continuous time limit, we get
\begin{equation}\label{eq:state2c}
\ket{\tilde \Psi (t)} =\!\! \sum_{k\in\{\mathrm{g,e}\}}\!\!\left(f^{(0)}_{j,k}(t)+\!\int_{0}^{t} dt^\prime f^{(1)}_{j,k}(t,t^\prime)e^{-i\omega_0 t^\prime} \au(t^{\prime})+\!\int_{0}^{t} dt^\prime \int_{t^\prime}^{t} dt^{\prime\prime} f^{(2)}_{j,k}(t,t^\prime,t^{\prime\prime})e^{-i\omega_0 (t^{\prime}\!+t^{\prime\prime})} \au(t^{\prime})\au(t^{\prime\prime})\right)\ket{k,\textbf{0}}
\end{equation}
and the coefficients $f^{(i)}_{j,k}$ are real-valued functions of time parametrised by $\Omega$ and $\gamma$, depending on the initial emitter state $\ket{j}$. Here, we truncate the state at the component with two photons emitted, since the amplitude of the $i$-photon component goes as $\gamma^{i/2}$ and the three-photon component is already negligible in our experimental situation.

Consider the case where the emitter is driven by $N$ coherent square pulses each with a duration $t_\mathrm{p}$ and with a separation between each pulse of $\Delta t_m$. As in the main text, we notate the pulses and time bins in reverse chronological order so that $\Delta t_1\rightarrow \infty$ is the period after the final pulse, allowing for the atom to return to its ground state. Then the total light-matter state can be obtained by combining the actions of the two evolution operators $U$ and $\tilde{U}$:
\begin{align}
\label{recursive}
\ket{\Psi}\equiv \ket{\text{g}}\ket{\psi_N}=U(\Delta t_{1})\tilde{U}(t_\mathrm{p})\cdots U(\Delta t_{N-1})\tilde{U}(t_\mathrm{p})U(\Delta t_{N})\tilde{U}(t_\mathrm{p})\ket{\mathrm{g},\textbf{0}},
\end{align}
where $\ket{\psi_N}$ is the photonic state produced after $N$ pulses. This expression takes into account the multi-photon emission for finite pulses and gives the time-dynamic expression for the photon wavepackets emitted by the atom within each time bin.

\subsection{A photonic Fibonacci sequence}
\label{subsupsec:fibo}

In the main text, we described how two sequential $\pi$-pulses separated by $T_{\!1/2}=\ln(2)T_1$ generates the maximally entangled Bell state. Here, we explore how a longer sequence of $\pi$-pulses can be used to generate multi-mode entangled states. An intuition for this process is given by first imagining that a sequence of $N$ ideal pulses far-separated in time will produce a time-bin product state $\ket{11\cdots 1}$ of $N$ photons, because the atom has time to decay between each pulse. If the time separations between the pulses are decreased, at some point, the atom may still be in the excited state when a subsequent pulse arrives. If so, this pulse will coherently bring the atom to the ground state, preventing the atom from emitting in the time bins preceding and following that pulse. This creates a correlated ``vacuum pair" within the stream of single photons $\ket{11\cdots 1001\cdots 1}$. The average number of vacuum states that are created increases as the time separations decrease but, because they always come in pairs, there are only certain ways that they can be organised within the stream of photons. For example, we can already see that only 5 possible photonic states can be produced by $N=4$ ideal $\pi$ pulses. These are the permutations: $\ket{0000}$, $\ket{0011}$, $\ket{1001}$, $\ket{1100}$, and $\ket{1111}$. If the light-matter interaction remains coherent over the timescale of the pulse sequence, the final photonic state must be a superposition of all possible permutations of vacuum pairs arranged among the stream of single photons.

Counting permutations of paired elements (00) among individual elements (1) arises in many different contexts, such as prosody in Greek and Indian poetry \cite{SINGH1985229Sup,knuth2013artSup}, where syllables of long length (vacuum pairs) are arranged between short ones (single photons). This pattern produces Pascal's triangle, the Fibonacci sequence, and the golden ratio, all of which arise in countless otherwise unrelated applications from phyllotaxis  \cite{mitchison1977phyllotaxisSup} to pulsating stars \cite{lindner2015strangeSup}. Notably, the recursive Fibonacci relation arises in quantum cryptography \cite{simon2013highSup} and quantum computing \cite{kauffman2018braidingSup}.

To reveal the relationship between the Fibonacci sequence and the photonic state produced by sequential $\pi$-pulses, we begin from Eq. (\ref{recursive}). If we assume that the pulses are much shorter than the lifetime of the atom, namely they are so short that we can neglect any emission during the pulse ($t_\mathrm{p}\ll1/\gamma$), then $\tilde{U}(t_\mathrm{p})$ becomes a unitary transformation on the atomic state only, and for a $\pi$-pulse we have $\tilde{U}(t_\mathrm{p})\rightarrow \hat{\sigma}_x$. In this case, at most one photon can be spontaneously emitted per time bin, and this emission is captured by the operator $U(\Delta t_m)$. This operator effectively performs a two-qubit gate between the atomic states $\{\ket{\text{g}},\ket{\text{e}}\}$ and the state of the $m$th time bin $\{\ket{0}_m,\ket{1}_m\}$, where $\ket{1}_m=\hat{t}_m^\dagger\ket{0}_m$. Drawing from Eq. (\ref{Eq:amplitudedamping}), in the ideal $\pi$-pulse scenario the $m$th time bin mode beginning at time $T_{[m]}$ can be defined using the pulse-mode formalism \cite{blow_continuum_1990Sup,hussain_quantum_1992Sup,ozdemir_pulse-mode_2002Sup} as $\hat{t}^\dagger_m=\int_0^\infty dt {f}_m^{(1)}(t)\hat{a}^\dagger(t)$, where the normalised complex one-photon wavefunction amplitude is ${f}_m^{(1)}(t)=\sqrt{\gamma}e^{-\gamma (t-T_{[m]})/2-i\omega_o (t-T_{[m]})}/\beta_m$ for $T_{[m]}\leq t\leq T_{[m]}+\Delta t_{m}$ and ${f}_m^{(1)}(t)=0$ otherwise. The normalization factor is given by $\beta_m^2=1-\alpha_m^2$ where $\alpha_m=e^{-\gamma\Delta t_m/2}$. This temporal second-quantised description of the propagating field is valid in the regime where the bandwidth of light is much narrower than its central frequency \cite{hussain_quantum_1992Sup}, which is the case for our experiments. See the next section for more details about partitioning time.

Since the photonic state $\ket{\psi_N}=\braket{\text{g}|\Psi}$ is generated by sequential interaction with the two-dimensional ancillary atom, it can be represented as a matrix product state (MPS) described by the set of $2\times 2$ isometries $V_{[m]}$ for $m\in\{1,\cdots,N\}$ acting on the initial atomic state $\ket{\text{g}}$ \cite{schon_sequential_2005Sup,schon_sequential_2007Sup}. Using the second-quantised time bin modes, we can determine $V_{[m]}$ by re-writing the spontaneous emission operator $U(\Delta t_m)$ as an amplitude damping map
\begin{equation}
    U(\Delta t_m):\left\{\begin{aligned}
    \ket{\text{g}}&\rightarrow\ket{0}_m\ket{\text{g}}\\
    \ket{\text{e}}&\rightarrow  \alpha_m\ket{0}_m\ket{\text{e}}+\beta_m\ket{1}_m\ket{\text{g}}
    \end{aligned}\right..
\end{equation}
Hence, the isometry $V_{[m]}=U(\Delta t_m)\hat{\sigma}_x$ is given by
\begin{equation}
    V_{[m]} = \alpha_m\ket{0}_m\ket{\text{e}}\bra{\text{g}}+\beta_m\ket{1}_m\ket{\text{g}}\bra{\text{g}}+\ket{0}_m\ket{\text{g}}\bra{\text{e}}
\end{equation}
Thus, the photonic state $\ket{\psi_N}$ is of the class MPS$_2$ and can be written as
\begin{equation}
    \ket{\psi_N} = \braket{\text{g}|V_{[1]}V_{[2]}\cdots V_{[N]}|\text{g}},
\end{equation}
where we again adopted the reverse-chronological labeling for convenience.

To reveal the Fibonacci-like recursive nature of the state, we can expand the MPS expression:
\begin{equation}
\label{fiborelation}
\begin{aligned}
    \ket{\psi_N}
    &=\alpha_N\ket{0}_N\braket{\text{g}|V_{[1]}V_{[2]}\cdots V_{[N-1]}|\text{e}}+\beta_N\ket{1}_N\braket{\text{g}|V_{[1]}V_{[2]}\cdots V_{[N-1]}|\text{g}}\\
    &=\alpha_N\ket{0}_N\ket{0}_{N-1}\ket{\psi_{N-2}}+\beta_N\ket{1}_{N}\ket{\psi_{N-1}}.
\end{aligned}
\end{equation}
This expression can be equivalently written in terms of the time bin mode operator $\hat{t}^\dagger_m$, as in the main text. We can now evaluate this expression to reveal the W-class entangled state produced by three pulses, knowing the first two terms are a single photon and a Bell-like entangled state:
\begin{equation}
\begin{aligned}
\ket{\psi_1}&=\ket{1}_1\\
  \ket{\psi_2}&=\alpha_2 \ket{0}_{2}\ket{0}_{1}+\beta_2 \ket{1}_{2}\ket{1}_1\\
  \ket{\psi_3}&=\alpha_3\ket{0}_{3}\ket{0}_2\ket{1}_1+\beta_3\alpha_2 \ket{1}_{3}\ket{0}_{2}\ket{0}_1+\beta_3\beta_2\ket{1}_{3}\ket{1}_2\ket{1}_1.
\end{aligned}
\end{equation}
By expanding the recursive wavefunction relation for a given number of pulses $N$, we can see that the product states composing $\ket{\psi_N}$ are given by all of the possible permutations of vacuum pairs ($00$) among single photons ($1$) arranged into the $N$ time bins, which matches the intuitive explanation. The number of unique product states $K_i$ containing $i$ vacuum pairs can be found as the sequence of numbers along the `shallow' diagonal of Pascal's triangle. The sum of this sequence is the Fibonacci number $F_N$ corresponding to the sequence generated by the recursive relation $F_N=F_{N-2}+F_{N-1}$ beginning  with $F_0=F_1=1$. Hence, $F_N=\sum_{i=0}^{\floor{N/2}}K_i$ is the total number of product states composing $\ket{\psi_N}$. This relationship can be directly identified from Eq. (\ref{fiborelation}) by noting that all product states with coefficient $\alpha_N$ must be distinct from those with coefficient $\beta_N$ due to having different states in their $N$th time bin. Since the number of product states composing $\ket{\psi_N}$ follows the Fibonacci sequence, it scales exponentially by $\varphi^N$ where $\varphi=(1+\sqrt{5})/2$ is the golden ratio.

Interestingly, there is a unique choice for the pulse separations $\Delta t_m$ so that all product state amplitudes are identical. This special condition is $\Delta t_{m\geq 2} = T_1\ln(F_{m}/F_{m-2})$, which provides the maximally entangled Bell ($N=2$) and W-class ($N=3$) states. It remains to study the amount and type of entanglement for states with $N\geq 4$. However, for large $N$ at this condition, we can identify that pulses near the beginning of a long sequence should have a constant `golden' pulse separation $\Delta t_m\rightarrow \Delta t_\varphi = 2T_1\ln(\varphi)$. As the sequence nears its end ($m\lesssim7$), the pulse separations should briefly fluctuate around $\Delta t_\varphi$ before ending with $\Delta t_3=T_1\ln(3)$ and $\Delta t_2=T_1\ln(2)$. Hence, the deviation of $\Delta t_m$ from $\Delta t_\varphi$ corrects for the truncation of the sequence.

\subsection{Time bin partitioning}

As described in the main text and in Refs. \cite{specht_single-atom_2011Sup,scweinthesisSup}, any pure single-photon wavepacket $\int dt f^{(1)}(t)\au(t)\ket{0}=\hat{t}^\dagger\ket{0}$ can be re-expressed as a single-photon entangled state $\alpha(T)\ket{01}_{el}+\beta(T)\ket{10}_{el}$ by partitioning it at the threshold time $T$. This is possible because we can always express the temporal mode as a sum of orthogonal modes $\hat{t}^\dagger=\alpha(T)\hat{t}^\dagger_e+\beta(T)\hat{t}_l^\dagger$, where $\alpha(T)\hat{t}^\dagger_e=\int_{-\infty}^T dt f^{(1)}(t)\au(t)$, $\beta(T)\hat{t}^\dagger_l=\int_T^\infty dtf^{(1)}(t)\au(t)$, $\alpha^2(T)=\int_{-\infty}^Tdt |f^{(1)}(t)|^2$, and $\beta^2(T)=\int_T^\infty dt|f^{(1)}(t)|^2$. Thus, we can define the amplitudes of the early and late time bin modes $f_e^{(1)}$ and $f^{(1)}_l$, respectively. This partitioning can be extended by defining $N-1$ threshold times $T_{[m]}$ so that the wavepacket is partitioned into $N$ orthogonal time bin modes. In this way, a single photon can also be re-expressed as an $N$-qubit entangled W state. For a single photon emitted by an exponentially-decaying atom with a lifetime $T_1$ the measured state will be a maximally entangled W state when choosing the threshold times to be $T_{[m]}{=}T_1\ln{\left(\frac{N}{N-m}\right)}$.

When two or more photons are present, it is not always possible to express the photonic state in terms of the orthogonal early and late modes. This is because a general two-photon amplitude $f^{(2)}(t_1,t_2)$ is not always separable into two single-photon amplitudes. In other words, the arrival time of the two photons could be entangled. However, in the case of the photonic sequence described in the previous section, only a single photon may be emitted between each of the short pulses and, if a single photon is emitted, its time of emission does not depend on the emission time of other photons, only on the atomic lifetime. Hence, by choosing $T_{[m]}$ to coincide with the pulse arrival times, the state can be re-expressed in terms of the second-quantised modes $\hat{t}^\dagger_m$, as was demonstrated in the previous sections.

Although we must choose to partition the wavepacket into time bin modes defined by the pulses, additional partitions could be made between pulses. For example, using the fact that any single-photon wavepacket can be partitioned into a $\ket{\psi^+}$ Bell state, we can re-express a $\ket{\phi^+}=(\ket{00}_{el}+\ket{11}_{el})/\sqrt{2}$ Bell state as $(\ket{000}_{eil}+(\ket{101}_{eil}+\ket{011}_{eil})/\sqrt{2})/\sqrt{2}$ W-class state by partitioning the early bin into two parts to create an intermediate ($i$) bin. Although this is one way to create multi-mode entangled states, the limitation is that partitioning a wavepacket into smaller and smaller bins makes it more difficult to separate, manipulate, and measure the modes with high fidelity.

\subsection{Beyond ideal pulses}
\label{subsupsec:N1}

We now go beyond the assumption that $t_\mathrm{p}\ll 1/\gamma$ to consider that emission during the pulse can take place. However, we assume that at most two photons are emitted in total. In this case, we can apply Eq. (\ref{recursive}) to find that
\begin{equation}
\label{eq:state_q_norm}
\ket{\psi}=\sqrt{p_0}\ket{\textbf{0}}+\sqrt{p_1}\int_{0}^{\infty}dt {f}^{(1)}(t)\au(t)\ket{\textbf{0}}+ \sqrt{p_2}\int_{0}^{\infty}\int_{0}^{\infty}dt_1 dt_2{f}^{(2)}(t_1,t_2)\au(t_2)\au(t_1)\ket{\textbf{0}}.
\end{equation}

\subsubsection{Single Pulse}
Emission from a two-level system after a single $\pi$ pulse has already been extensively studied \cite{fischer2018scatteringSup,scweinthesisSup}. In this case, the ${f}^{(2)}$ component is primarily composed of one photon emitted during the excitation pulse of width $t_\mathrm{p}$ and one photon emitted afterwards at the rate $\gamma$. This source of error is referred to as re-excitation noise and it is one of two factors causing a non-zero $g^{(2)}$---the second being imperfect suppression of the excitation pulse. In general, the two photons composing ${f}^{(2)}$ are temporally entangled due to the sequential nature of their emission from a two-level atom. However, for a pulse much shorter than $1/\gamma$, the two photons are approximately separable ${f}^{(2)}(t_1,t_2)\simeq {f}^{(1)}_\mathrm{n}(t_1){f}^{(1)}(t_2)$, where $t_1\leq t_2$ and ${f}^{(1)}_\mathrm{n}$ is the temporal wavefunction of the noise photon that has a very small overlap with the single-photon component \cite{ollivier_hong-ou-mandel_2021Sup,scweinthesisSup}. In this fast-pulse regime, ${f}^{(1)}$ and ${f}^{(1)}_\mathrm{n}$ are given by \cite{scweinthesisSup}
\begin{equation}
\label{singlepulsewavefunction}
\begin{aligned}
    {f}^{(1)}(t)\simeq\frac{\sqrt{\gamma}e^{-i\omega_o t}}{\sqrt{p_1}}
    \left\{
    \begin{aligned}
    &\sin\left(\Omega t/2\right)\cos\left(\Omega(t_\mathrm{p}-t)/2\right)e^{-\gamma t_\mathrm{p}/4}&t<t_\mathrm{p}\\
    &\sin\left(\Omega t_\mathrm{p}/2\right)e^{-\gamma\left(2t-t_\mathrm{p}\right)/4}&t\geq t_\mathrm{p}
    \end{aligned}
    \right.\\
    {f}^{(1)}_\mathrm{n}(t)\simeq\frac{\sqrt{p_1\gamma}e^{-i\omega_o t}}{\sqrt{p_2}}
    \left\{
    \begin{aligned}
    &\sin\left(\Omega t/2\right)\csc\left(\Omega t_\mathrm{p}/2\right)\sin\left(\Omega(t_\mathrm{p}-t)/2\right)&t<t_\mathrm{p}\\
    &0&t\geq t_\mathrm{p}
    \end{aligned}
    \right.\\
\end{aligned}
\end{equation}
The values of $p_0$, $p_1$, and $p_2$ can then be obtained from the normalisation conditions of $\ket{
\psi}$, ${f}^{(1)}$, and ${f}^{(2)}$.

\subsubsection{Two Pulses}
For sequential $\pi$ pulses, the temporal wavefunctions ${f}^{(1)}$ and ${f}^{(2)}$ can be decomposed into the four orthogonal intervals $\Delta T_1=(0,t_\mathrm{p})$, $\Delta T_2=(t_\mathrm{p},\Delta t)$, $\Delta T_3=(\Delta t,\Delta t+t_\mathrm{p})$, $\Delta T_4=(\Delta t+t_\mathrm{p},\infty)$. This decomposition provides wavefunction expressions similar in nature to Eq.~(\ref{singlepulsewavefunction}) but where ${f}^{(1)}$ is a piecewise sum of 4 wavefunctions and ${f}^{(2)}$ a piecewise sum of 10 wavefunctions, corresponding to the 10 ways of placing 2 photons among the 4 intervals.

To simplify the discussion, we introduce the probabilities $P_{n_1,n_2,n_3,n_4}$, where $n_i$ are the number of photons emitted during time interval $\Delta T_i$. The one-photon probability is then $p_1 = P_{1000}+P_{0100}+P_{0010}+P_{0001}$. In the same way, the two-photon probability is $p_2\simeq P_{1100}+P_{1010}+P_{0110}+P_{1001}+P_{0101}+P_{0011}$, where the cases $P_{0200}=P_{0002}=0$ vanish because two-photon emission can never occur while the pulse is off and we neglect the very small contribution from $P_{2000}$ and $P_{0020}$, which can only occur during the short pulse intervals.

To define our final two time bins, we group the intervals $\Delta T_1$ through $\Delta T_3$ into an early bin $\Delta T_e=(0,T)$ where $T$ here is fixed to be $T\equiv \Delta t +t_\mathrm{p}$, followed by the late bin $\Delta T_l=\Delta T_4=(T,\infty)$. This allows us to consider ${f}^{(1)}$ as the sum of two orthogonal parts: ${f}_{10}(t)$, where the photon is emitted during the early bin, and $F_{01}(t)$, where it is emitted during the late bin. Their associated probabilities are $p_{10}= P_{1000}+ P_{0100}+P_{0010}$ and $p_{01}=P_{0001}$, respectively. In the same way, ${f}^{(2)}$ becomes the sum of two orthogonal parts: ${f}_{20}(t_1,t_2)$, where the two photons are both emitted during $\Delta T_e$, and ${f}_{11}(t_1,t_2)$, where one photon is emitted during $\Delta T_e$ and the other in $\Delta T_l$. Their associated probabilities are $p_{20}= P_{1100}+P_{1010}+P_{0110}$ and $p_{11}=P_{1001}+P_{0101}+P_{0011}$. With this choice of time bin basis, the wavefunction ${f}_{11}(t_1,t_2)$ can be always factored as ${f}_{11}(t_1,t_2)={f}_e(t_1){f}_l(t_2)$ for a truncated early photon ${f}_e(t_1)$ followed by an exponentially-decaying late photon ${f}_l(t_2)=\sqrt{\gamma}e^{-\gamma t_2/2-i\omega_o t_2}$. Using all these temporal wavefunction components, we can generate theory predictions for all measured quantities, including fidelity and concurrence.

The photonic state re-expressed in this time bin basis reads
\begin{equation}
\begin{aligned}
\label{eq:state_q_norm_2}
\ket{\psi_2}&=\sqrt{p_{0}}\ket{00}+\sqrt{p_{01}}\ket{01}+\sqrt{p_{10}}\ket{10}+ \sqrt{p_{20}}\ket{20}+\sqrt{p_{11}}\ket{11}
\end{aligned}
\end{equation}
where $\ket{02}$ states cannot occur by definition of the chosen partition and we have, for example, $\ket{10}=\int_{0}^{T}\!\!dt {f}_{10}(t)\au(t)\ket{\mathbf{0}}$. From this model in the case of short $\pi$-pulses, we have
\begin{equation}
\label{eq0011prbs}
\begin{aligned}
p_{0} &\simeq e^{-\gamma\Delta t}\\
p_{01}&\simeq 0\\
p_{10}&\simeq \frac{1}{4}\gamma t_\mathrm{p} e^{-\gamma\Delta t}\\
p_{20}&\simeq \frac{3}{8}\gamma t_\mathrm{p}\left(e^{-\gamma t_\mathrm{p}}-e^{-\gamma\Delta t}\right)\\
p_{11} &\simeq\left(\frac{3}{8}\gamma  t_\mathrm{p}+1\right)e^{-\gamma t_\mathrm{p}}+ \left(\frac{3}{8}\gamma t_\mathrm{p}-1\right)e^{-\gamma\Delta t}.
\end{aligned}
\end{equation}
The above probabilities do not sum to $1$ due to the truncation of the basis. Because of this, we can estimate the three photon emission probability by taking $p_3\simeq 1-(p_0+p_1+p_2)$ where $p_1=p_{01}+p_{10}$ and $p_{2}=p_{20}+p_{11}$. In addition, the intrinsic overlap between the first and second emitted photons of the two-photon component is estimated given by the ratio $p_{20}/p_2\simeq p_{20}/p_{11}$. For short pulses, this reduces to a linear dependence on pulse width: $\sim 3\gamma t_\mathrm{p}/8$.

\section{Theoretical model of the measurement}
\label{supsection:theorymeasurement}

To model the measurements of coherence, we use the approach detailed in section 3.2.3 of Ref.~\cite{scweinthesisSup}, which is an extension of that presented in Ref.~\cite{kiraz2004quantumSup}. Under the assumption that the imperfect photonic states arriving at the final balanced beam splitter of the Mach-Zehnder interferometer are uncorrelated and identical, but for the interferometer phase $\phi$, the coincidences of the detectors monitoring the outputs are given in terms of the input field correlations by
\begin{equation}
    2G_\text{HOM}^{(2)}(t_1,t_2)=
    N(t_1)N(t_2)-\left|G^{(1)}(t_1,t_2)\right|^2+G^{(2)}(t_1,t_2)-\left|{C}^{(2)}(t_1,t_2)\right|^2\cos(2\phi)+2{C}^-(t_1,t_2)\cos(\phi),
\end{equation}
where $N(t)=\braket{\au(t)\ad(t)}$ is the input intensity, $G^{(2)}(t_1,t_2)=\braket{\au(t_1)\au(t_2)\ad(t_2)\ad(t_1)}$ is the intensity correlation, $G^{(1)}(t_1,t_2)=\braket{\au(t_2)\ad(t_1)}$ characterises the temporal purity of the state, and ${C}^{(2)}(t_1,t_2)=\braket{\ad(t_2)\ad(t_1)}$ captures the second-order coherence. The antisymmetric term ${C}^-(t_1,t_2){=}\text{Re}\!\left[\braket{\ad(t_2)}\!\braket{\au(t_2)\au(t_1)\ad(t_1)}\!{-}\!\braket{\ad(t_1)}\!\braket{\au(t_1)\au(t_2)\ad(t_2)}\right]$ is a modification of the Hong-Ou-Mandel interference dynamics in the presence of first-order coherence. For two-level atoms, this term can only be nonzero during coherent driving. Hence, its contribution to $G^{(2)}_\text{HOM}$ is negligible for the cases studied in this work where we only apply short $\pi$ pulses.

To explore the properties of the photonic states when subdivided into two time bins, we integrate $G^{(2)}_\text{HOM}$ over the four quadrants defined by the threshold $T$ separating early $e$ and late $l$ time bins. Then, to obtain quantities that are loss independent, we normalise by the squared average photon number for each pair of bins. This gives
\begin{equation}
\label{eq:g2homrelation}
    2g^{(2)}_{\text{HOM},ab} = 1-M_{ab}+g^{(2)}_{ab}-{c}^{(2)}_{ab}\cos(2\phi)+2{c}^-_{ab}\cos(\phi),
\end{equation}
where the subdivided quantities for $a,b\in\{e,l\}$ are defined by
\begin{equation}
\label{eq:bipartite}
\begin{aligned}
    g^{(2)}_{ab}&=\frac{1}{\mu_a\mu_b}\int_a\int_b G^{(2)}(t_1,t_2)dt_1 dt_2&\hspace{20mm}
    M_{ab}&=\frac{1}{\mu_a\mu_b}\int_a\int_b\left|G^{(1)}(t_1,t_2)\right|^2dt_1 dt_2\\
    {c}^{(2)}_{ab}&=\frac{1}{\mu_a\mu_b}\int_a\int_b\left|{C}^{(2)}(t_1,t_2)\right|^2dt_1 dt_2&\hspace{20mm}
    {c}^-_{ab}&=\frac{1}{\mu_a\mu_b}\int_a\int_b{C}^-(t_1,t_2)dt_1 dt_2\\
\end{aligned}
\end{equation}
and $\mu_a=\int_a N(t)dt$. Note that, due to the time symmetry of $G^{(2)}(t_1,t_2)$, $\left|G^{(1)}(t_1,t_2)\right|^2$, and $\left|{C}^{(2)}(t_1,t_2)\right|^2$, we have that $g^{(2)}_{ab}=g^{(2)}_{ba}$, $M_{ab}=M_{ba}$, and ${c}^{(2)}_{ab}={c}^{(2)}_{ba}$, respectively. However, ${C}^-(t_1,t_2)$ is antisymmetric and so ${c}^{-}_{ee}={c}^{-}_{ll}=0$ and ${c}_{el}^-=-{c}_{le}^-$. Eq.~(1) of the main text can then be recovered by taking the weighted average
\begin{equation}
    g^{(2)}_\text{HOM} = \overline{\mu}_e^2g^{(2)}_{\text{HOM},ee}+\overline{\mu}_e\overline{\mu}_l\left(g^{(2)}_{\text{HOM},el}+g^{(2)}_{\text{HOM},le}\right)+\overline{\mu}^2_lg^{(2)}_{\text{HOM},ll},
\end{equation}
where $\overline{\mu}_a=\mu_a/\mu$ is the proportion of intensity in bin $a\in\{e,l\}$ and $\mu=\mu_e+\mu_l=\int N(t)dt$.

If the input has first-order coherence ($\braket{\ad(t)}\neq 0$), the detector rates fluctuate in opposition by $\mu^{\pm}=\mu[1\pm {c}^{(1)}\cos(\phi)]$ where ${c}^{(1)}=\mu^{-1}\!\int\left|\braket{\ad(t)}\right|^2dt$ is the integrated squared magnitude of the first-order coherence \cite{loredo_generation_2019Sup}. This oscillation gives the self-homodyne signal $I_\text{SH}$ discussed in the main text. Like $g^{(2)}$, we normalise $g^{(2)}_\text{HOM}$ with respect to the uncorrelated coincident counts obtained for detection delays $|t_1-t_2|$ greater than the laser repetition period. However, these counts are also affected by the interferometer phase if $I_\text{SH}\neq 0$ \cite{gustin2018pulsedSup,scweinthesisSup}. The average coincident counts between uncorrelated detection events is the product of the average photon number received at each detector $\mu^+\mu^-=\mu^2(1-I_\text{SH}^2)$, which can underestimate $\mu^2$. This underestimate remains even after phase averaging: $I_\text{SH}^2\rightarrow ({c}^{(1)})^2/2$. For the cases studied in this work, ${c}^{(1)}$ is measured to be less than 0.1 for pulse separations above $39$~ps implying a $0.5\%$ normalisation error. For the half-life case of $\Delta t=98$~ps, ${c}^{(1)}\simeq 0.03$ giving an expected error of $0.05\%$. Hence, the uncorrelated coincidence counts give a good approximation of $\mu^2$ and the normalisation for each quadrant $\mu_a\mu_b$.

\subsection{Density matrix elements and fidelity estimates}
\label{supsubsec:fidelityestimates}

In general, the measured state is not a pure state as in section \ref{supsection:theorymodel}, but a mixed state described by a photonic density matrix $\hat{\rho}$. This reduction in purity can be caused by decoherence processes such as electron-phonon interactions of the quantum dot emitter. We can decompose this density matrix by photon number: $\hat{\rho}=\sum_{m,n}\sqrt{p_mp_n}\hat{\rho}_{m,n}$, where $\text{Tr}[\hat{\rho}_{n,n}]=1$ and $\hat{\rho}_{n,m}=\hat{\rho}_{m,n}^\dagger$. The form of $\hat{\rho}$ can be obtained from the form of $\ket{\psi}\bra{\psi}$ given in Eq.~(\ref{eq:state_q_norm}), with the exception that the normalised temporal density functions $\xi^{(m,n)}$ cannot, in general, be factored into amplitudes ${f}^{(m)}{f}^{(n)*}$. For example, the imperfect single-photon state in the time basis is given by $\hat{\rho}_{1,1}=\iint\xi^{(1)}(t,t^\prime)\hat{a}^\dagger(t)\ket{0}\bra{0}\hat{a}(t^\prime)dtdt^\prime$, where we have used the simplified notation $\xi^{(n)}=\xi^{(n,n)}$. To make an explicit connection to the model presented in section~\ref{supsection:theorymodel}, the single photon emitted by a purely dephased atom can be approximated by $\xi^{(1)}(t,t^\prime)\simeq {f}^{(1)}(t){f}^{(1)*}(t^\prime)e^{-\gamma^\star|t-t^\prime|}$, where $\gamma^\star\ll\Omega$ is the pure dephasing rate such that $\gamma+2\gamma^\star$ is the FWHM of the homogeneously-broadened photon intensity spectrum \cite{scweinthesisSup}.

To describe fidelity, we first define the second-quantised time bin modes as $\hat{t}_a^\dagger=\int_a f_a(t)\au(t)dt$ where $\int_a|f_a(t)|^2dt=1$ for $a\in\{e,l\}$. Then, the density matrix elements of the photonic state in this basis are $\varrho_{klmn}=\braket{kl|\hat{\rho}|mn}$ where $\ket{mn}=(\hat{t}_e^\dagger)^m(\hat{t}_l^\dagger)^n\ket{\mathbf{0}}/\sqrt{m!n!}$. These elements involve overlap integrals between the ideal temporal wavefunctions and the temporal density functions $\xi^{(m,n)}$. However, all the quantities measured with the self-homodyne setup correspond to purity measurements where the overlap integrals are between $\xi^{(m,n)}$ exclusively.

Purity measurements can give a good estimate of the magnitude of some density matrix elements $|\varrho_{klmn}|$. Using the single-photon case as an illustrative example, by the Cauchy-Schwarz inequality we have
\begin{equation}
\label{supeq:csi}
\begin{aligned}
    \left|\braket{\mathbf{0}|\hat{t}_a\hat{\rho}\hat{t}^\dagger_b|\mathbf{0}}\right|^2 &=p_1^2\left|\int_a dt_a\int_bdt_b\iint dt dt^\prime f_a^*(t_a)\xi^{(1)}(t,t^\prime)f_b(t_b)\text{Tr}[\ad(t_a)\au(t)\ket{0}\bra{0}\ad(t^\prime)\au(t_b)]\right|^2\\
    &=p_1^2\left|\int_a dt\int_b dt^\prime (f_a^*(t)f_b(t^\prime))\xi^{(1)}(t,t^\prime)\right|^2\lesssim p_1^2\int_a dt\int_b dt^\prime\left|\xi^{(1)}(t,t^\prime)\right|^2,
\end{aligned}
\end{equation}
where we have used the fact that $f_a$ and $f_b$ are normalised. This inequality is saturated when the photon can be written in the time-bin basis: $\xi^{(1)}(t,t^\prime)\propto f_a(t)f_b^*(t^\prime)$ for $t$ in $a$ and $t^\prime$ in $b$, which can only be satisfied for a pure photon. However, by defining our modes to have the most optimal shape $|f_{a}(t)|^2=\xi^{(1)}(t,t)$ (for $t$ in $a$) and phase that best matches $\xi^{(1)}(t,t^\prime)$, this upper-bound approximation is very accurate when dephasing is small; for instance, when the single-photon indistinguishability (trace purity) $M_\mathrm{s}=\text{Tr}[\hat{\rho}_{1,1}^2]=\iint|\xi^{(1)}(t,t^\prime)|^2dtdt^\prime$ from the source is greater than 0.5 (see sections 3.3.3 and 3.3.4 of Ref.~\cite{scweinthesisSup}), which is the case for the device studied here.

\begin{figure}
    \centering
    \includegraphics[width=0.6\textwidth]{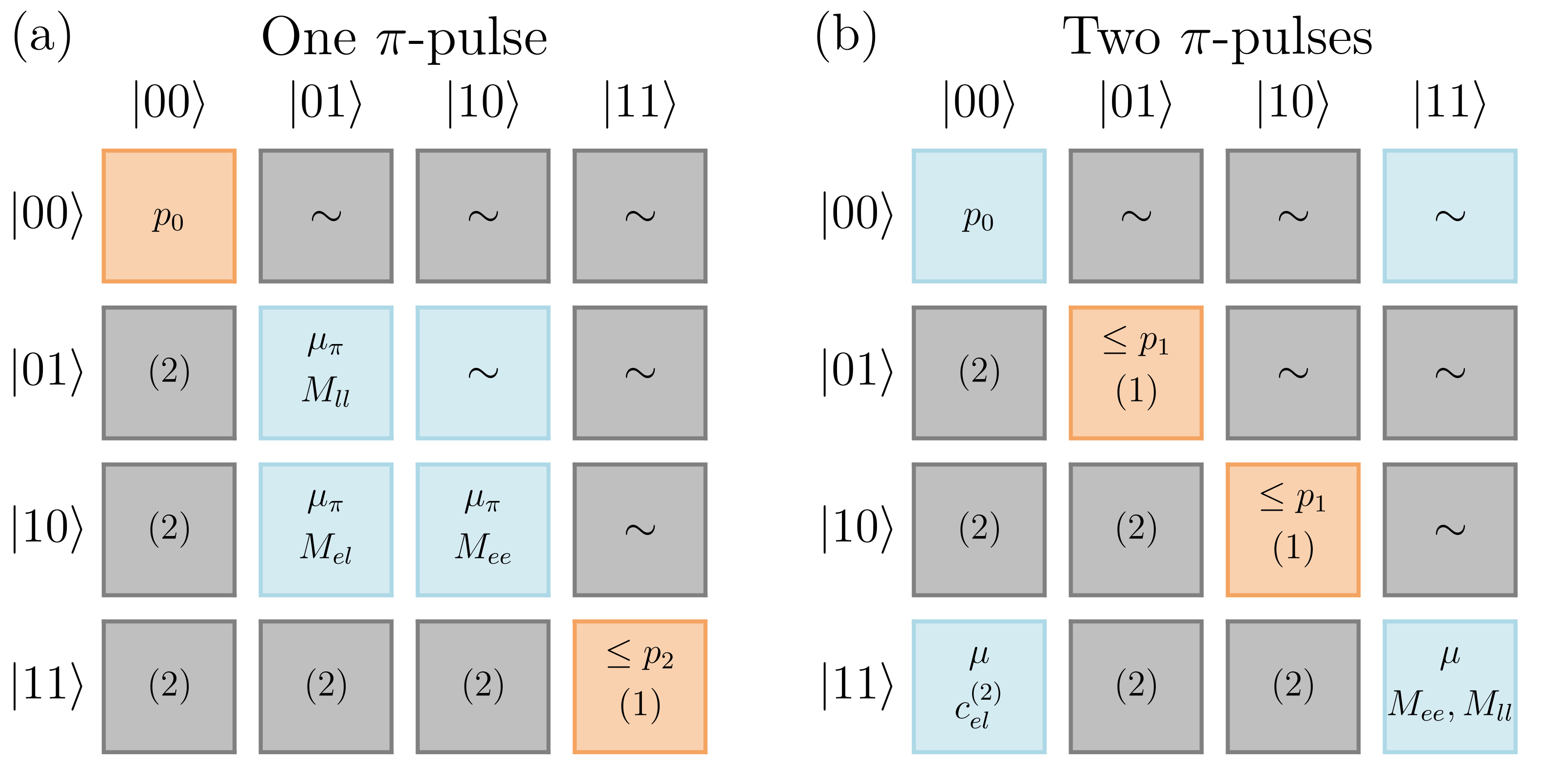}
    \caption{\textbf{Summary diagram of estimations of the photonic density matrix elements} after (a) one $\pi$ pulse and (b) two $\pi$ pulses, corresponding to the $\ket{\psi^+}$ type entangled state and the $\ket{\phi^+}$ type entangled state, respectively. The blue coloured squares indicate the elements estimated using intensity measurements ($\mu_\pi$ or $\mu$) and coherence measurements ($M_{ab}$ or $c^{(2)}_{el}$) that are required to estimate the Bell-state fidelity. The orange squares indicate the unknown diagonal elements bounded by the photon number probabilities $p_n$ determined from intensity correlation measurements $g^{(2)}$ and $g^{(3)}$. The gray squares indicate the unknown off-diagonal elements and the symbol $\sim$ indicates that the element is determined from the Hermitian property. The numbers (1) and (2) indicate the number of free parameters. Note that the important elements illustrated by the blue squares are estimated under different assumptions, namely that $p_1\gg p_2$ after one pulse and $p_0p_2\gg p_1p_3$ after two pulses.}
    \label{fig:densitymatrixestimates}
\end{figure}

We would like to now estimate the right-hand-side of Eq.~(\ref{supeq:csi}) using measurements of the mean wavepacket overlaps $M_{ab}$. However, any measurement of $\xi^{(1)}$ after losses will include contribution from the multi-photon `noise' components $\xi^{(n\geq 2)}$ if $g^{(2)}\neq 0$. In other words, $M_{ab}$ quantifies $G^{(1)}=p_1\xi^{(1)}+\mathcal{O}(p_2)$ (recall Eq.~\ref{eq:bipartite}). We can identify two extreme cases: (i) $G^{(1)}\simeq p_1\xi^{(1)}$, implying that the noise contributes nothing to the purity measurement (e.g. $\mu_\pi^2M\sim p_1^2M_\mathrm{s}$), and (ii) $G^{(1)}\simeq\mu_\pi\xi^{(1)}$ implying that the noise contributes proportionally to the single-photon subspace (e.g. $M\sim M_\mathrm{s}$). Taking the former will overestimate the fidelity in most cases whereas the latter will generally underestimate the fidelity. In principle, one could have $G^{(2)}>\mu_\pi\xi^{(1)}$, which implies that the presence of noise purifies the total state (e.g. $M>M_\mathrm{s}$), but this is extremely unlikely given that the same dephasing processes are expected to degrade all subspaces. To take the possible range from case (i) to (ii) into account, we estimate $\left|\braket{\mathbf{0}|\hat{t}_a\hat{\rho}\hat{t}^\dagger_b|\mathbf{0}}\right|^2\simeq \tilde{\mu}_\pi^2\overline{\mu}_a\overline{\mu}_b M_{ab}$ where $p_1\leq\tilde{\mu}_\pi\leq \mu_\pi$. This gives a fidelity estimate of
\begin{equation}
\label{supeq:fidpsiplus}
    \mathcal{F}_{\psi^+}=\frac{1}{2}\left(\varrho_{0101}+\varrho_{1010}+\varrho_{0110}+\varrho_{1001}\right)\simeq\frac{\tilde{\mu}_\pi}{2}\left(\overline{\mu}_e\sqrt{M_{ee}}+\overline{\mu}_l\sqrt{M_{ll}}+2\sqrt{\overline{\mu}_e\overline{\mu}_l M_{el}}\right),
\end{equation}
where $\overline{\mu}_a=\mu_a/\mu_\pi$ and where we have defined the difference in phase of $t_e^\dagger$ and $t_l^\dagger$ such that $\varrho_{1001}=\varrho_{0110}\geq 0$. The photonic density matrix structure after one $\pi$ pulse is summarized in Fig.~\ref{fig:densitymatrixestimates}~(a).

The fidelity $\mathcal{F}_{\psi^+}$ is bounded from above by $\tilde{\mu}_\pi\sqrt{M}$ using the generalised mean inequality. In fact, from Eq.~(\ref{supeq:csi}) we can find that it is more precisely bounded by $p_1\sqrt{M_\mathrm{s}}$. For the device used in this work, $M_\mathrm{s}$ can be accurately estimated by $M_\mathrm{s}=M/(1-g^{(2)})$ \cite{ollivier_hong-ou-mandel_2021Sup}, which corresponds to a scenario half-way between cases (i) and (ii) discussed above.

We apply this same approach to estimate the fidelity for the two-pulse case, where we wish to compute the fidelity to the Bell state $\ket{\phi^+}=(\ket{00}+\ket{11})/\sqrt{2}$. The diagonal density matrix element corresponding to $\ket{00}$ is simply given by the vacuum probability $\varrho_{0000}=p_0$. The important coherence element can be estimated from the second-order coherence $c^{(2)}_{el}$ between early and late time bins similar to $M_{el}$ for the single-photon entangled state: $|\varrho_{0011}|^2=|\varrho_{1100}|^2=\tilde{\mu}^2\overline{\mu}_e\overline{\mu}_lc^{(2)}_{el}$. However, to compute the diagonal element associated with $\ket{11}$ from our measurements, we must assume that the two-photon density function is approximately separable in arrival time $\xi^{(2)}(t_1,t_2,t_1^\prime,t_2^\prime)\simeq\xi^{(1)}_1(t_1,t_1^\prime)\xi^{(1)}_2(t_2,t_2^\prime)$. That is, we assume that the arrival times of the two photons are not entangled, although they may still overlap in time and be impure. This is a very reasonable assumption for our experiment given that the only moment two-photon temporal entanglement can be created is during the pulses, which are brief compared to the total wavepacket timescale. If this assumption is not satisfied, and subsequent measurements on the same photonic state are made, it could lead to a state description better suited for pseudo-density matrices that can have negative eigenvalues [19]. However, all the density matrices in our analysis are assumed a priori---and verified a posteriori---to be positive semi-definite Hermitian matrices.

Knowing the important four density matrix elements, we estimate the Bell-state fidelity with respect to $\ket{\phi^+}$ by
\begin{equation}
\label{supeq:phiplusfidelity}
    \mathcal{F}_{\phi^+}=\frac{1}{2}\left(\varrho_{0000}+\varrho_{1111}+\varrho_{0011}+\varrho_{1100}\right)\simeq\frac{1}{2}\left(p_0+\frac{\tilde{\mu}^2}{p_2}\overline{\mu}_e\overline{\mu}_l\sqrt{M_{ee}M_{ll}}+2\tilde{\mu}\sqrt{\overline{\mu}_e\overline{\mu}_l {c}^{(2)}_{el}}\right),
\end{equation}
where $2p_2\leq\tilde{\mu}\leq\mu$ and where we have defined the sum of the phases of $\hat{t}_e^\dagger$ and $\hat{t}_l^\dagger$ so that $\rho_{0011}=\rho_{1100}\geq 0$. The photonic density matrix structure after two $\pi$ pulses is summarized in Fig.~\ref{fig:densitymatrixestimates}~(b). The presence of $p_2^{-1}$ in Eq. (\ref{supeq:phiplusfidelity}) arises due to the fact that $\braket{11|\hat{\rho}|11}\propto p_2$ whereas $M_{ee}M_{ll}\propto p_2^4$. One can see that in the case where $\tilde{\mu}=2p_2$, $\overline{\mu}_e=\overline{\mu}_l=1/2$, we find the more intuitive term $p_2\sqrt{M_{ee}M_{ll}}$. Note that $M_{ee}$, $M_{ll}$, and ${c}^{(2)}_{el}$ are already directly degraded by the detector jitter. Hence, the intensity overlap does not explicitly arise in the fidelity estimate.

\subsection{Concurrence estimates}
\label{supsubsec:concurrenceestimates}

A high state fidelity gives an indication of how close the photonic state is to being a photon-number Bell state. However, the fidelity does not indicate if the state is indeed an entangled state. A nonzero entanglement concurrence \cite{wootters2001entanglementSup} $0< \mathcal{C}\leq 1$ unequivocally indicates that a state is entangled. Computing the concurrence requires full knowledge of the two-qubit density matrix, which cannot be obtained without performing a full quantum state tomography. However, as described in the previous section, we can still estimate the four most important density matrix elements for the single- and two- $\pi$-pulse cases based on both intensity and phase correlation measurements. Then, a range for all unknown elements can be estimated by assuming the matrix must be positive semi-definite and Hermitian.

To estimate the entanglement concurrence, we build the two-qubit density matrix with two free parameters for each unknown off-diagonal matrix element and one free parameter for each unknown diagonal element, such as $\varrho_{1101}=c_{1101}e^{i\phi_{1101}}$, where $0\leq c_{1101}\leq 1$ and $0\leq\phi_{1101}\leq 2\pi$. In addition, we constrain the diagonal elements to not exceed the corresponding measured photon number probabilities $p_n$. These constructed density matrices are summarized in Fig.~\ref{fig:densitymatrixestimates}. We then sample the free parameters from a uniform distribution within their allowed range and sample the measured values from a normal distribution with a standard deviation given by the measurement uncertainty. We repeat sampling while rejecting unphysical density matrices until we obtain $10^5$ matrices that are positive semi-definite. We then take the mean concurrence of the random sample to be the estimated concurrence corresponding to our measurements. This analysis approach is applied to the $\ket{\psi^+}$ case generated after a single $\pi$ pulse in section \ref{Ssection:analysispi} and to the $\ket{\phi^+}$ case generated after the double $\pi$ pulse sequence in section \ref{Ssection:analysis2pi}.

\vspace{-2mm}
\section{Experimental analysis}
\label{supsection:experimental}

\begin{figure*}
    \centering
    \includegraphics[width=\textwidth,trim=0 2mm 0 0, clip]{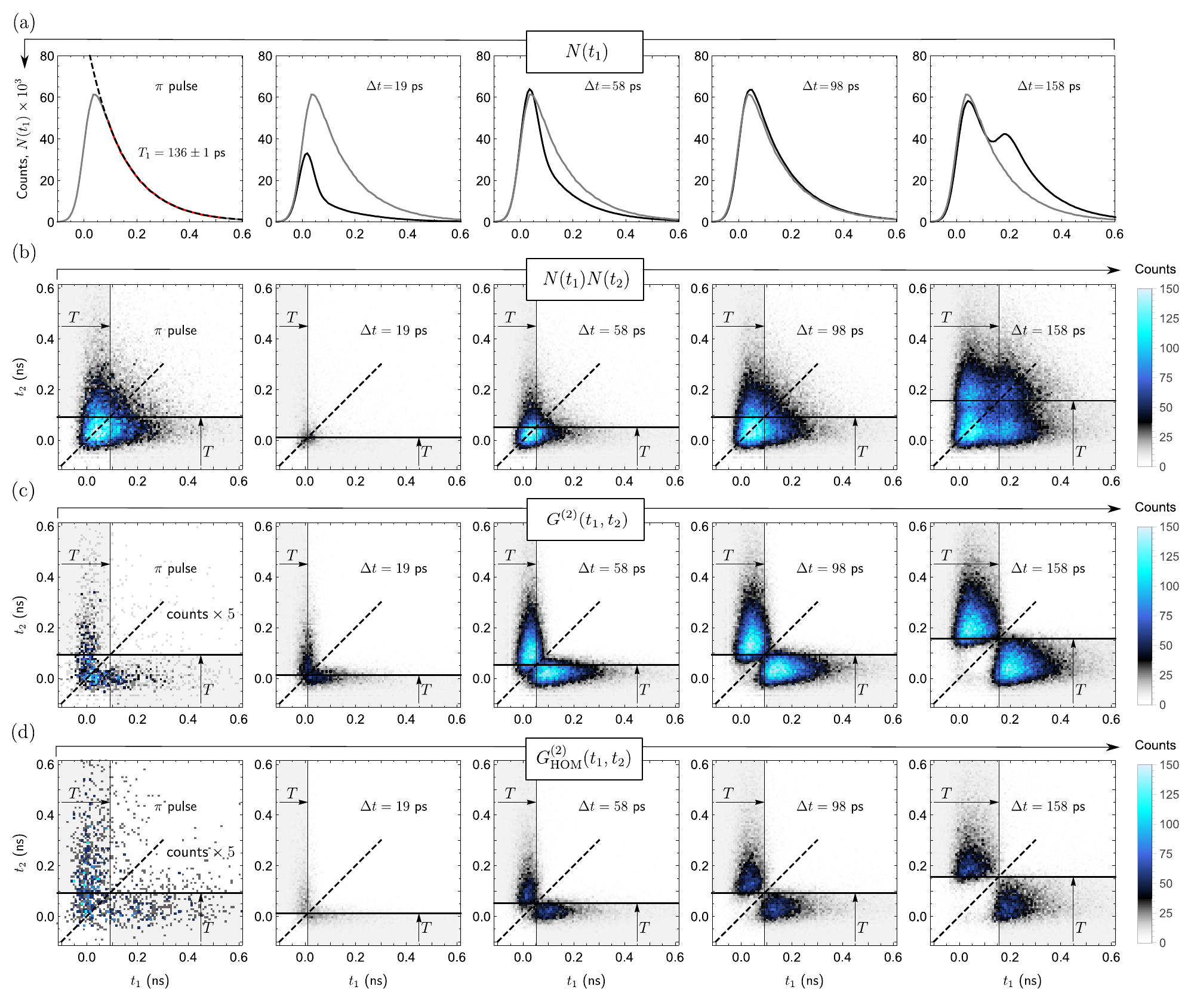}\vspace{-4mm}
    \caption{\textbf{Time-resolved photon intensity and correlation measurements.} The rows display, from top to bottom, (a) the intensity $N(t_1)$ from direct measurement (grey/black full trace corresponds to single/double $\pi$-pulse excitation), (b) the two-time intensity $N(t_1)N(t_2)$ obtained from the uncorrelated counts of the Hanbury Brown-Twiss (HBT) setup, (c) the intensity correlation $G^{(2)}(t_1,t_2)$ from the correlated counts of the HBT setup, and (d) the intensity correlation $G^{(2)}_\text{HOM}$ after Hong-Ou-Mandel (HOM) like interference at the end of the path-unbalanced Mach-Zehnder interferometer. The columns display, from left to right, the $\pi$-pulse case and four selected two-pulse cases with pulse separations $\Delta t= 19$~ps, 58~ps, 98~ps, and 158~ps. All panels have a time resolution of 8~ps, which for the two-time maps is $8\times 8$~ps$^2$. The time axes units for all panels are nanoseconds. The dashed trace in panel (a) is a mono-exponential fit to the  temporal profile decay.}
    \label{Sfigure:brick}
\end{figure*}

\vspace{-3mm}
\subsection{Time-resolved measurements}
\vspace{-3mm}

In this section, we present the full set of time-resolved measurements after applying a single $\pi$ pulse as well as for several selected delays $\Delta t$ between sequential $\pi$ pulses. Fig.~\ref{Sfigure:brick}~(a) shows the intensity profiles corresponding to $N(t)$ as well as the deduced atomic lifetime of $T_1=136\pm1$~ps from the $\pi$-pulse case. We also show the two-time map $N(t_1)N(t_2)$ (Fig.~\ref{Sfigure:brick}~(b)) obtained from the uncorrelated counts of the Hanbury Brown-Twiss setup, which are used to normalise $G^{(2)}(t_1,t_2)$ (Fig.~\ref{Sfigure:brick}~(c)). Finally, Fig.~\ref{Sfigure:brick}~(d) shows the two-time intensity correlation at the output of the Mach-Zehnder interferometer $G^{(2)}_\text{HOM}(t_1,t_2,\phi)$ after summing over all measured cases of $\phi$. For the two-time maps corresponding to a single pulse, we show quadrants divided by $T$ chosen at the half-life time. For maps corresponding to sequential pulses, the quadrants are defined by choosing $T$ to coincide with the arrival of the second pulse. 

The temporal overlap of the two photons of the $\alpha\ket{00}+\beta\ket{11}$ state relies on the temporal characteristics of the excitation pulses and the detection jitter time. These two imperfections cause a bleeding of the counts from the off-diagonal quadrants into the diagonal quadrants of the maps shown in Fig.~\ref{Sfigure:brick}~(c) and (d) for the double $\pi$ pulse cases. As discussed in the \textit{Methods} section, our detectors present a Gaussian jitter time of $s\simeq50$ ps FWHM. The excitation laser pulses are $ t_\mathrm{p}\simeq 20$ ps long FWHM. From these values, we measure that there is an $\sim 18\%$ temporal overlap between the two photons, which degrades the Bell state fidelity. Using the estimate $3\gamma t_\mathrm{p}/8$ from theory, we expect only a 5\% overlap due to the pulse width of 20 ps. Since both the pulse width and the detector jitter have the similar effect of blurring the wavepacket via a convolution, we can estimate that the total measured overlap goes like $(3\gamma/8)\sqrt{ t_\mathrm{p}^2+s^2}$ for $ t_\mathrm{p},s < 1/\gamma$. This simple estimate predicts an overlap of 15\% based on our experimental parameters, which is not far from the measured overlap of 18\%.

\begin{figure*}
    \centering
    \includegraphics[width=1\textwidth]{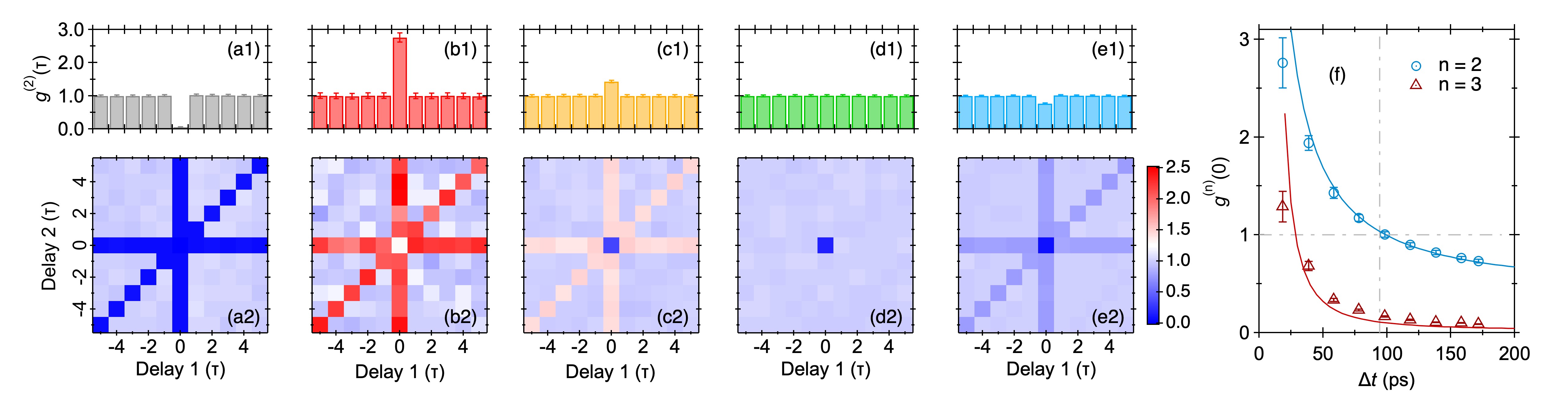}\vspace{-3mm}
    \caption{\textbf{Normalised second and three-photon correlation maps} measured after (a) single $\pi$-pulse excitation and (b-e) double $\pi$-pulse excitation for the same $\Delta t$ delays indicated in Fig. S2. The first/second row compiles the normalised two-/ three-photon coincidence histogram $g^{(2)}(\tau)$/$g^{(3)}(\tau_1,\tau_2)$. The counts in the first row are normalized by the average of the uncorrelated counts which are, from panel (a1) to (e1): 76100, 25500, 201200, 366400, and 1207700, respectively.  (f) Corresponding $g^{(n)}(0)$ values, $n{=}2,3$, for all studied delays. The solid curves show the theory predictions. The $(\tau)$ units in the axis of panels (a-e) are equal to 12.3 ns. The error bars in the first row (a1 through e1) and panel (f) indicate the standard deviation obtained assuming Poissonian statistics in the detected events within each peak.
    }
    \label{Sfigure:G3maps}
\end{figure*}

We measure the integrated intensity correlations $g^{(2)}(\tau)$ and $g^{(3)}(\tau_1,\tau_2)$ by passing the emission through a set of two cascaded fiber beam splitters (with 1:3 and 1:1 splitting ratios, respectively). The three output fibers are connected to three detectors $i=1,2,3$ clicking at time $t_i$, where triple-detection events are recorded as a function of the delays $\tau_1=t_1-t_2$ and $\tau_2=t_2-t_3$. We then extract the two-photon correlation $g^{(2)}(\tau)$ from either detector pair 1-2, 1-3 or 2-3. In Fig.~\ref{Sfigure:G3maps}, we show the normalised time-integrated two and three photon correlations as a function of the detector photon arrival delay times. The measurements after a single $\pi$-pulse are given in panels (a) and for each pulse separation for double $\pi$-pulse excitation in panels (b) through (e).

For a given delay 1 and 2, each point in the $g^{(3)}(\tau_1,\tau_2)$ map corresponds to the integrated (and normalised) triple-coincidences in a temporal delay square of $5{\times}5$ ns$^2$ area. The simultaneous triple coincidences, $g^{(3)}(0)\simeq 6p_3/\mu^3$, are located in the centre of the map. Additionally, the simultaneous double coincidences, $g^{(2)}(0)\simeq (2p_2{+}6p_3)/\mu^2$, are located all along the vertical, horizontal and positive diagonal directions. The non-normalised three-photon coincidence histograms, used to extract the normalised $g^{(3)}(\tau_1,\tau_2)$ histograms shown in the third row of Fig. \ref{Sfigure:G3maps}, are not included in this supplementary, but they are available in the permanent link 10.6084/m9.figshare.16838248.

Different phenomenology can be observed in this set of measurements. The $\pi$-pulse excitation (panels (a)) presents a strong antibunching for both two and three simultaneous coincidences, highlighting the predominance of single photon emission, where $g^{(2)}(0){=}0.063\pm0.002$ and $g^{(3)}(0){=}0.0016{\pm}0.0009$. In the series of different delays compiled in panels (b-e) and summarized in panel (h), we observe a gradual decrease of $g^{(3)}(0)$ and $g^{(2)}(0)$ with increasing separation between the two pulses. The most remarkable feature is the change of the photon statistics for $g^{(2)}(0)$ from bunching when $\Delta t < T_1$ to antibunching when $g^{(2)}(0) >T_1$ and $g^{(2)}(0)\simeq 1$ when $\Delta t\simeq T_1$.

\vspace{-2mm}
\subsection{Photon probabilities and loss}
\label{supsection:probabilitiesandloss}
\vspace{-2mm}

\begin{figure*}
    \centering
    \includegraphics[width=0.95\textwidth,trim=1 1 1 1, clip]{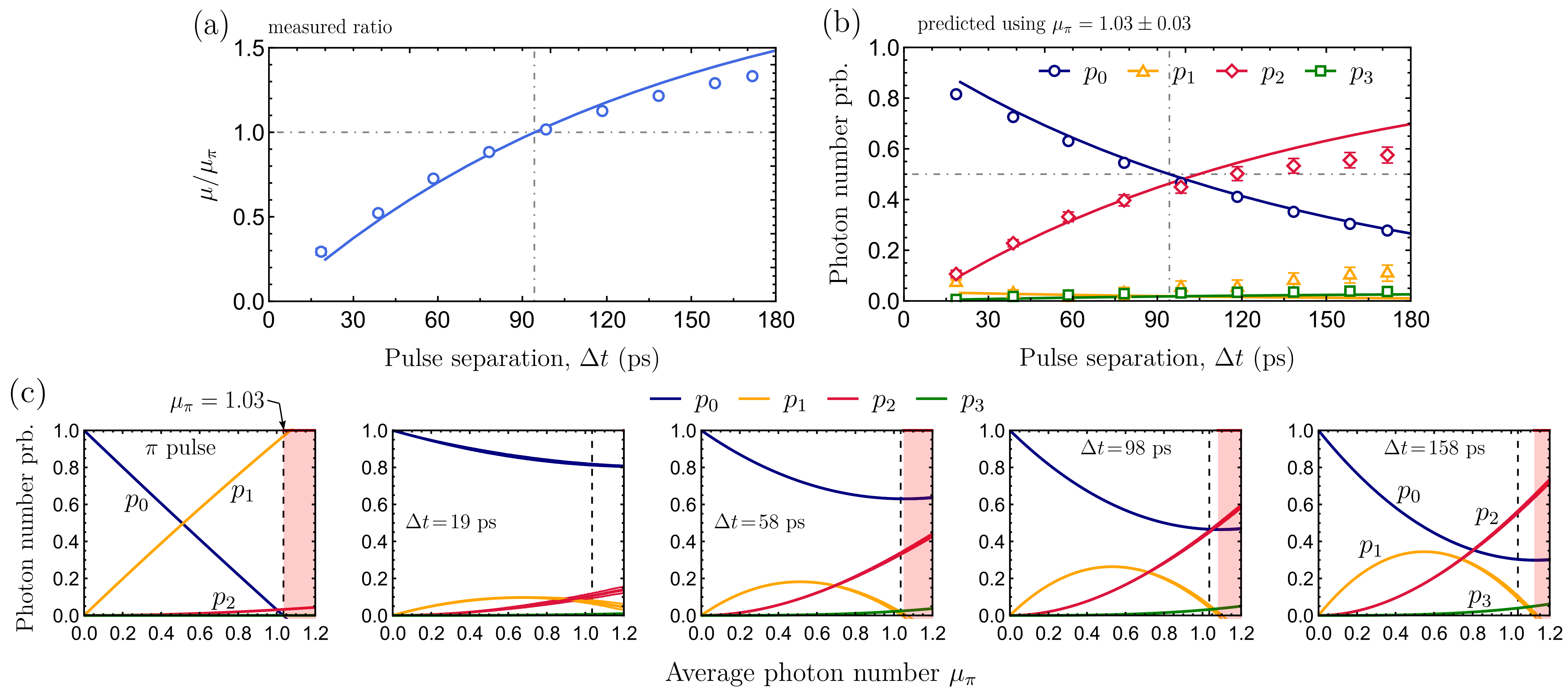}\vspace{-3mm}
    \caption{\textbf{Photon number probabilities.} (a) The measured total average photon number $\mu/\mu_\pi$ for two $\pi$ pulses separated by delay $\Delta t$ relative to $\mu_\pi$ detected after a single $\pi$ pulse. (b) The probabilities $p_n$ for the source to emit $n$ photons given $\mu_\pi=1.03\pm0.03$. The solid lines in panels (a) and (b) show the theory predictions. (c) The range of experimentally estimated photon probabilities for losses parameterised by $\mu_\pi$. The red shaded region shows the unphysical regime where at least one $p_n$ becomes negative. The vertical dashed line shows the value of $\mu_\pi$ such that $p_0$ is nearly zero for the $\pi$-pulse case. It is also the largest value of $\mu_\pi$ such at all probabilities extracted from the measurements are physical. Error bars in panels (a) and (b), as well as the error bands in panel (c) indicate the standard deviation obtained from propagating the Poissonian statistics of the single-photon counting measurements. Note that the error bars in panel (a) are smaller than the points.}
    \label{Sfigure:pns}
\end{figure*}

All measurements described in section \ref{supsection:theorymeasurement} are normalised, hence independent of photon losses, aside from the average photon number $\mu$. However, fidelity is loss-dependent in general. Thus, to estimate the fidelity and photonic state probabilities at source, we must estimate $\mu$ at the source. For the double $\pi$ pulse cases, we measure $\mu$ after losses and normalise it with respect to the value $\mu_\pi$ detected for a single $\pi$ pulse after the same amount of losses (see Fig.~\ref{Sfigure:pns}~(a), a similar measurement is performed in the work of Liu and coworkers~\cite{liu_high_2018Sup}, where the double $\pi$-pulse excitation is used to extract $T_1$). Thus, all measurements can be projected back to the source provided an estimate of $\mu_\pi$ at the source.

To estimate $\mu_\pi$ at the source, we make only one assumption: the probability $p_0$ that no photons are emitted is very small when applying one $\pi$ pulse excitation. This is a very reasonable assumption, provided that the quantum dot is in the correct electronic configuration. That is, the photon probabilities and fidelity estimated in this work are corrected for losses as well as any potential source blinking.

Using the measured values of $\mu/\mu_\pi$, $g^{(2)}$, and $g^{(3)}$, we can calculate the photon probabilities for all cases studied as a function of $\mu_\pi$ (see Fig.~\ref{Sfigure:pns}~(c)). In this figure, we see that the value of $\mu_\pi=1.03\pm0.03$ is the most reasonable choice to estimate probabilities at the source. Fig.~\ref{Sfigure:pns}~(b) shows the photon number probabilities at the source predicted from the measurements using $\mu_\pi=1.03\pm0.03$, where we find good agreement with theory. The estimated $p_n$ values after a single $\pi$ pulse are found to be $p_0=0.01\pm0.03$, $p_1=0.96\pm0.03$, $p_2=0.035\pm0.003$, and $p_3=0.0003\pm0.0002$. For the highlighted two $\pi$ pulse case where $\Delta t=98$~ps$\simeq T_{\!1\!/2}$, we find that $p_0=0.47\pm0.01$, $p_1=0.05\pm0.04$, $p_2=0.45\pm0.03$, and $p_3=0.032\pm0.003$. These probabilities are crucial to estimate the Bell state fidelity.

In our experiments, we chose the source displaying the highest brightness of those characterized in Ref. \cite{Ollivier:2020aaSup}, with a single-photon brightness of $18\%$ at the first lens, and a single photon rate in the single-mode collection fiber of 8.9 MHz (with a laser repetition rate 81.08 MHz). This allowed for fast and efficient acquisition of photon correlation events. The first lens brightness is dependent on four main factors: (1) A Purcell factor of about 7 implies $90\%$ of the QD emission goes into the cavity mode. (2) The asymmetrical microcavity design yields a cavity mode emission efficiency towards the first collection of $\sim80\%$ \cite{hilaire_deterministic_2020Sup,Ollivier:2020aaSup}. (3) The cross-polarised filtering technique used to remove the excitation laser pulses halves the collected single-photon emission from the charged QD emission. (4) The efficiency of the charge load in the trion QD state \cite{hilaire_deterministic_2020Sup}, which impacts the excitation efficiency of the charged QD.

Further steps can be taken towards the deterministic operation of the entanglement source by placing the first lens inside the cryostat chamber and using optimal coatings in the optical elements in the collection setup. Optimising the tunneling barrier of the device can improve the charge load efficiency and its stabilisation.

\vspace{-4mm}
\subsection{Time bin analysis of a single $\pi$ pulse}
\label{Ssection:analysispi}
\vspace{-2mm}

\begin{figure*}
    \centering
    \includegraphics[width=0.98\textwidth]{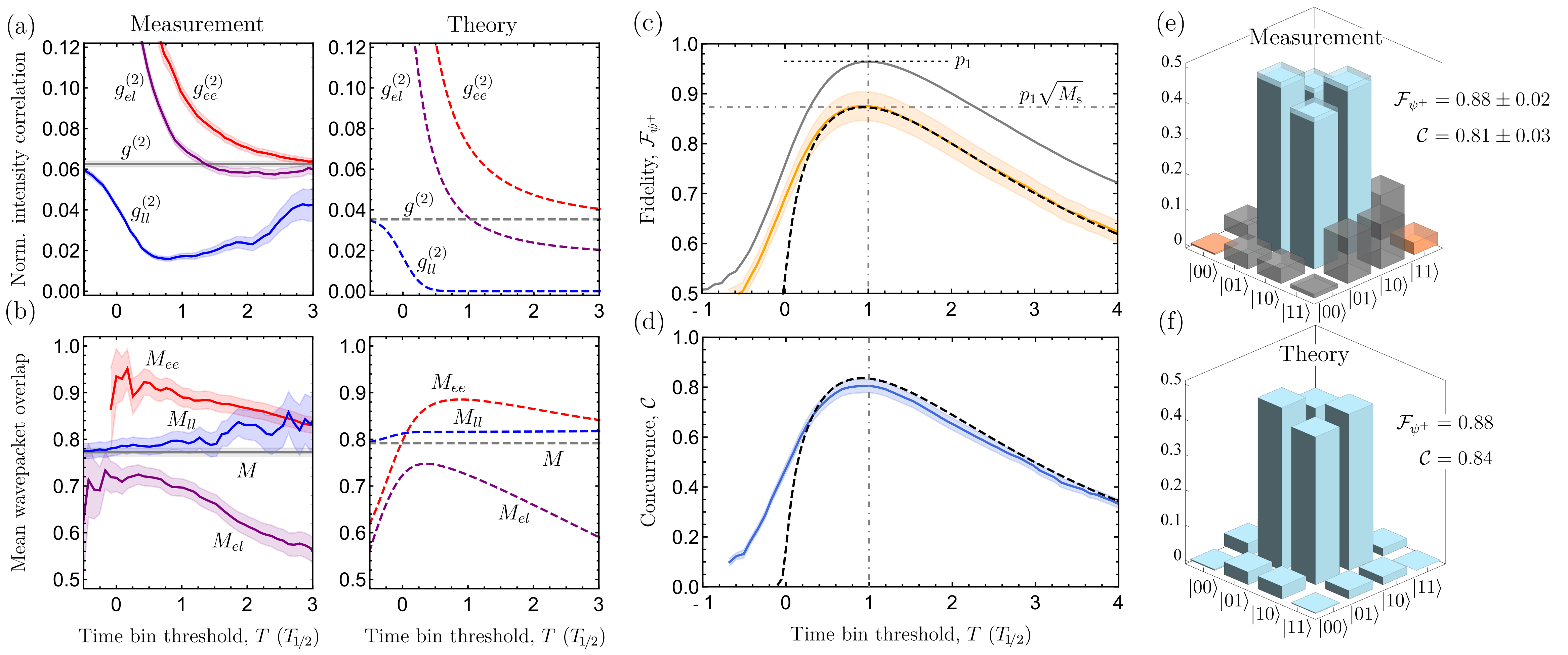}\vspace{-3mm}
    \caption{\textbf{Time bin analysis of a single photon.} The measured values of the (a) subdivided normalised intensity correlation $g^{(2)}_{ab}$ along with the total integrated value $g^{(2)}$ and (b) subdivided mean wavepacket overlaps $M_{ab}$ for $a,b\in\{e,l\}$ along with the total integrated value $M$ as a function of time bin threshold $T$ in units of the half-life $T_{\!1\!/2}$. Here, we compute $g^{(2)}_{el}=g^{(2)}_{le}$ and $M_{el}=M_{le}$ by averaging the off-diagonal quadrants of two-time correlation maps. The error bands indicate the standard deviation given by the Poissonian statistics of the single-photon counts within each time bin. As opposed to showing the theory for the ideal state like in the main text, which would here give $M_{ab}=M=1$ and $g^{(2)}_{ab}=g^{(2)}=0$, we show the theory prediction that includes dephasing and multi-photon emission. For these panels only, we also apply a convolution to account for a 50~ps detector jitter. (c) The Bell-state fidelity (orange) using Eq.~(\ref{supeq:fidpsiplus}) along with the upper bound (gray, solid) from intensity measurements alone. The shaded orange region indicates the range of fidelity estimated by taking $\tilde{\mu}_\pi$ to be $0.96=p_1\leq \tilde{\mu}_\pi\leq\mu_\pi=1.03$. (d) Estimated entanglement concurrence as a function of the time bin threshold. The connected blue line indicates the mean value of the concurrence and the shaded region indicates the standard deviation from the mean estimated by randomly sampling $10^5$ possible physical density matrices that correspond to our measurements for each value of $T$. (e) Partial density matrix reconstruction using experimental data at $T=T_{\!1\!/2}$ corresponding to the vertical line in panels (c) and (d). Blue elements are estimated from time-resolved intensity and coherence measurements directly. The gray elements represent the upper bound on the magnitude of the unknown off-diagonal elements, knowing the corresponding diagonal elements. The orange elements are estimated from the photon-number probabilities. The translucent region at the top of the blue bars indicate the standard deviation. (f) The density matrix computed using the theoretical model. All theory predictions (dashed curves and the density matrix) are computed for the state after applying pure dephasing to the wavefunctions given by Eq.~(\ref{singlepulsewavefunction}) to obtain $\xi^{(1)}$ as described in section \ref{supsubsec:fidelityestimates}. We use the parameters $1/\gamma=136$~ps, $\Omega=\pi/ t_\mathrm{p}$ for $ t_\mathrm{p}=20$~ps, and a pure-dephasing rate of $\gamma^\star=0.11\gamma$ corresponding to $M_\text{s}=\gamma/(\gamma+2\gamma^\star)=0.82$.
    }
    \label{Sfigure:psiplusbellstate}
    \vspace{-3mm}
\end{figure*}

The single-particle entanglement carried by a pure photon can be revealed by splitting it in half in a given degree of freedom. For example, splitting a photon in half using a balanced beam splitter gives rise to a photon-number $\ket{\psi^+}$ type Bell state encoded in path \cite{van2005singleSup}. Here, we split the photon in half by resolving the wavepacket in time. 

The two factors that reduce the quality of the single-photon Bell state are temporal dephasing and a multi-photon component arising from the imperfect excitation pulse. These are quantified by the mean wavepacket overlap $M$ and intensity correlation $g^{(2)}$, respectively. By subdividing the $G_\text{HOM}(t_1,t_2)$ and $G(t_1,t_2)$ maps measured after a $\pi$ pulse into four quadrants (recall the first column of Fig.~\ref{Sfigure:brick}), as done in the main text, we acquire the subdivided quantities $M_{ab}$ and $g^{(2)}_{ab}$ for a chosen threshold time $T$. This gives a detailed picture of the single-photon quality.

The subdivided intensity correlations $g^{(2)}_{ab}$, presented in Fig.~\ref{Sfigure:psiplusbellstate}~(a), are generally larger than than predicted by our theory (see Fig.~\ref{Sfigure:psiplusbellstate}~(a)), although the trends match. We expect that this is additional noise perhaps arising from imperfect suppression of the excitation laser, which is not accounted for in our model. However, the fact that $g^{(2)}_{ll}$ remains high for $T$ much larger than the pulse width suggests that a constant background noise may instead be the culprit that explains the extra $\sim\!0.02$ to the total $g^{(2)}$.

When sweeping $T$ across the wavepacket, we also find that $M_{ee}$, $M_{ll}$, and $M_{el}$ remain quite high (see Fig.~\ref{Sfigure:psiplusbellstate}~(b)), as opposed to the dip observed for $M_{el}$ in the main text. This is desired for the single-photon $\ket{\psi^+}$ Bell state as it attests to the indistinguishability of photons in each bin (high $M_{ee}$ and $M_{ll}$) in addition to the large amount of coherence between the states $\ket{10}$ and $\ket{01}$ (high $M_{el}$). Note that, a large $M_{el}$ in this case does not imply that the photons in the early and late bins overlap in time. Such a conclusion could only be drawn if the early and late time bin modes were not entangled. We can also again note that $M_{ee}$ is larger than the total value $M$ due to temporal truncation, which is also predicted by our model. The sharp drop in $M_{ee}$ and $M_{el}$ predicted as $T$ approaches the pulse is due to the sharp rise in $g^{(2)}_{ee}$ and $g^{(2)}_{el}$ from re-excitation events. Because the intensity is low in this region, our measurements have a large uncertainty but may still hint at a similar dropping trend. We also find, in both experiment and theory, that $M_{ll}$ is slightly larger than $M$ due to the fact that it is less affected by multi-photon events, which should primarily occur in the early bin.

Using Eq.~(\ref{supeq:fidpsiplus}), we find that the Bell-state fidelity is maximum at the half-life $T=T_{\!1\!/2}$, as expected (see Fig.~\ref{Sfigure:psiplusbellstate}~(c)). This maximum occurs when $\overline{\mu}_e=\overline{\mu}_l\simeq 1/2$ so that the photon is split perfectly in half. However, not only does $p_1\simeq0.96<1$ reduced the fidelity, but the reduced purity in time also affects the measured fidelity. We find that the measured $M_{ab}$ along with Eq.~(\ref{supeq:fidpsiplus}) agree well with the maximum fidelity predicted by $p_1\sqrt{M_\text{s}}$ where $M_\text{s}\simeq0.82$.

Lastly, we give an estimate of the concurrence\cite{wootters2001entanglementSup} $\mathcal{C}$, where a value $\mathcal{C}>0$ indicates entanglement (see Fig.~\ref{Sfigure:psiplusbellstate}~(d)). Using the approach described in section \ref{supsubsec:concurrenceestimates}, we uniformly sample $10^5$ physical density matrices from the partially reconstructed density matrix for each value of $T$ while also accounting for the normally distributed measurement uncertainty. An example of the partially reconstructed density matrix for $T=T_{\!1\!/2}$ is shown in Fig.~\ref{Sfigure:psiplusbellstate}~(e). From this analysis, we estimate a maximum concurrence of $\mathcal{C}=0.81\pm0.03$ at the Bell-state condition $T=T_{\!1\!/2}$, which corresponds very well to the theoretical prediction of $\mathcal{C}=0.84$ as shown in Fig.~\ref{Sfigure:psiplusbellstate}~(f). We find that the uncertainty from the normally-distributed measured quantities (blue bars in Fig.~\ref{Sfigure:psiplusbellstate}~(e)) dominate the uncertainty in the estimated concurrence rather than the uniformly distributed free parameters (dark gray bars in Fig.~\ref{Sfigure:psiplusbellstate}~(e)). Thus, the concurrences of the $10^5$ sampled matrices are normally distributed about their mean. Therefore, we take the standard deviation of the sample to represent the uncertainty of the estimated concurrence.

For theory calculations in panels (c), (d), and (f), we compute the density matrix elements of the imperfect photonic state $\xi^{(1)}(t,t^\prime)={f}^{(1)}(t)F^{(1)*}(t^\prime)e^{-\gamma^\star|t-t^\prime|}$ relative to the temporal modes ${f}_e$ and ${f}_l$ defined by normalising the pure temporal wavefunction $f_a(t)\propto {f}^{(1)}(t)$ for $t$ in bin $a$. However, since our model underestimates $p_2$ by about a factor of 2, we have used the measured values of $p_1$ and $p_2$ along with the normalised wavefunctions from the model. Note that, unlike the fidelity, our theory slightly overestimates the concurrence. This is because the theory wavefunctions underestimate $\varrho_{1111}$ compared to our measurements that indicate $g^{(2)}_{ll}\neq 0$ even for large $T$. This increased $\varrho_{1111}$ element degrades the concurrence but not fidelity.

\vspace{-2mm}
\subsection{Time bin analysis for two $\pi$ pulses with variable separation in time}
\label{Ssection:analysis2pi}
\vspace{-2mm}

\begin{figure*}
    \centering
    \includegraphics[width=0.63\textwidth]{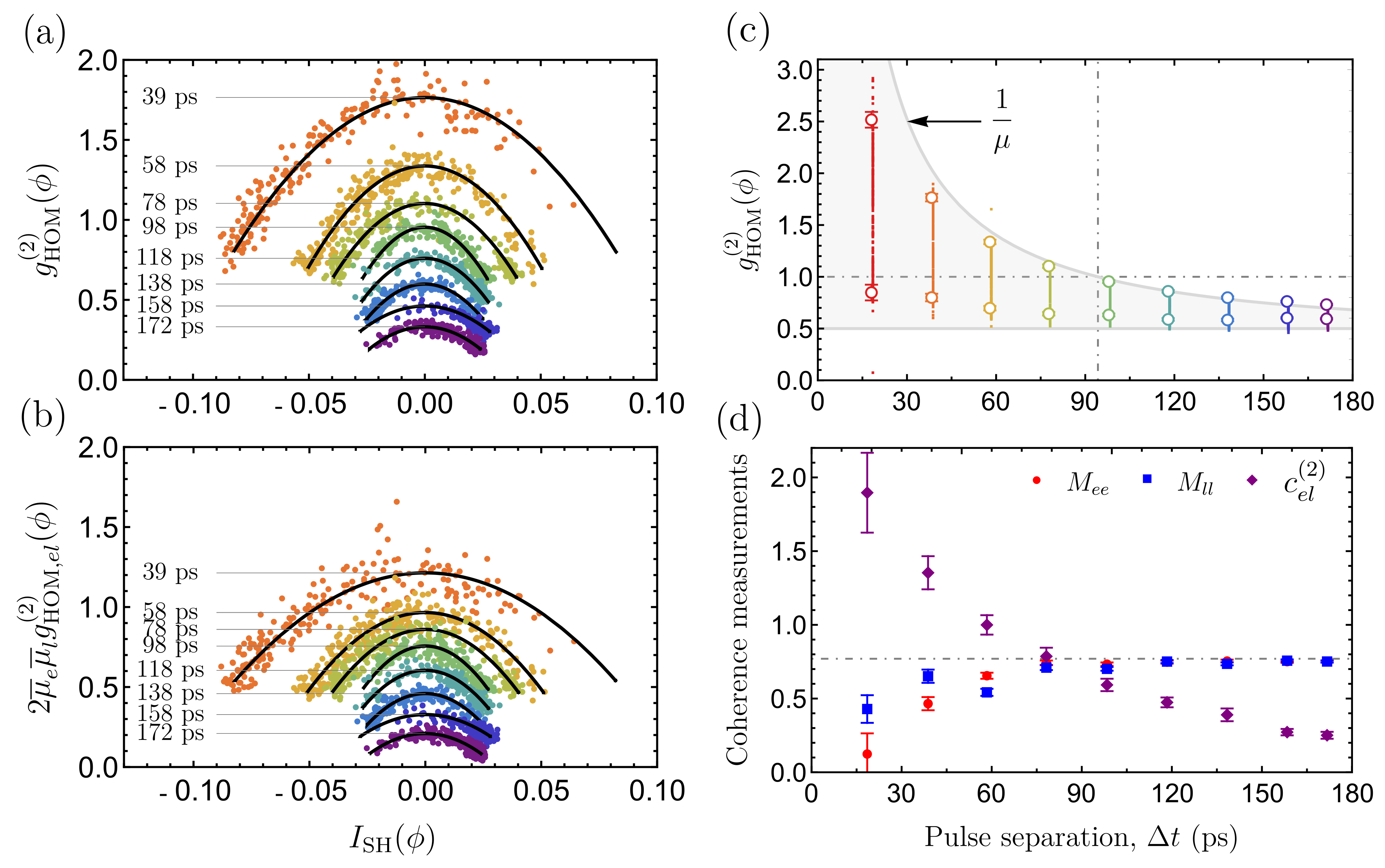}\vspace{-3mm}
    \caption{\textbf{Dependence of phase correlation measurements on the pulse separation time.} (a) The observed phase-correlated data showing the fitted quadratic curves (black, solid). The data set for $\Delta t=19$~ps is neglected because its range of oscillation is very large. (b) The observed signatures after excluding all coincident counts occurring in the diagonal quadrants of $G^{(2)}_\text{HOM}(t_1,t_2,\phi)$, when the time-bin threshold is chosen to coincide with the arrival of the second pulse. For clarity, in panels (a) and (b), we have shifted the data sets for $\Delta t=$ 118 ps, 138 ps, 158 ps, and 172 ps downwards by 0.1, 0.2, 0.3, and 0.4, respectively. (c) The range of $g^{(2)}_\text{HOM}$ as a function of the pulse separation time where the colored markers indicate the minimum and maximums from the fits in panel (a). The gray region bounded by $1/2\leq g^{(2)}_\text{HOM}\leq\mu^{-1}$ shows the theoretically expected range of oscillation for an ideal two-pulse experiment. The dashed gray lines indicate the maximum $g^{(2)}_\text{HOM}=1$ expected for the half-life case $\Delta t=T_1\ln(2)$. (d) Extracted measurements of mean wavepacket overlap and second-order coherence for each pulse separation. The horizontal line indicates the measured single-photon mean wavepacket overlap $M\simeq0.77$. Error bars in panels (c) and (d) indicate the standard error of the regression from fitting the scattered data.
    }
    \label{Sfigure:rainbowfigure}
    \vspace{-3mm}
\end{figure*}

 \begin{figure*}
    \centering
    \includegraphics[width=0.95\textwidth]{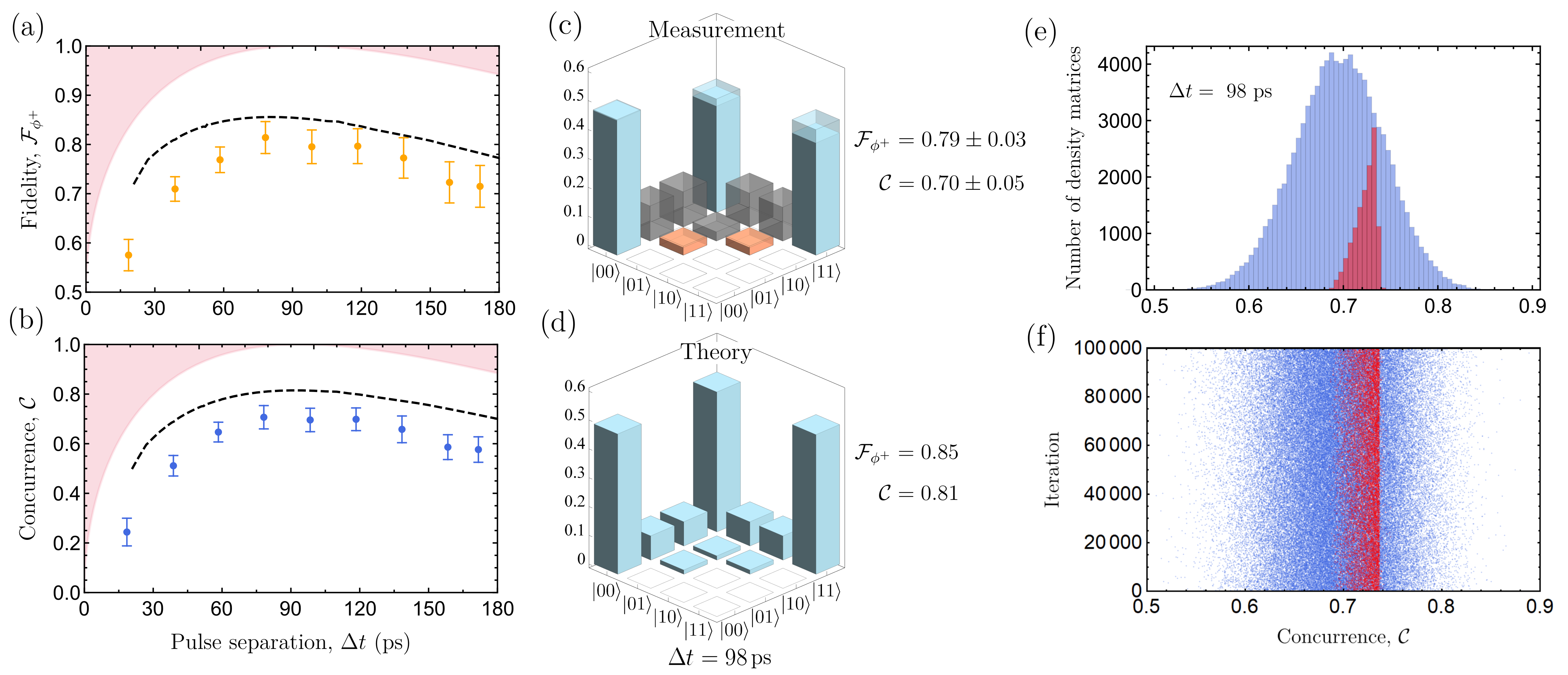}\vspace{-3mm}
    \caption{\textbf{Fidelity and concurrence.} (a) Bell-state fidelity and (b) concurrence (blue dots) estimated from measurements of second-order coherence and the photon-number probabilities. The error bars indicate the standard deviation. The red shaded area in both panels indicates the inaccessible region for an ideal two-pulse experiment. The black dashed curve shows the theory predictions obtained by applying pure dephasing to the wavefunctions given by the model in section \ref{supsection:theorymodel}, using parameters $\Omega=\pi/ t_\mathrm{p}$, $ t_\mathrm{p}=20$~ps, $1/\gamma=136$~ps, and $\gamma^\star=0.11\gamma$. Note that these curves do not include degradation due to detector jitter and hence over-estimate the experimental values. (c) Partial reconstructed density matrix using experimental data for pulses separated by $\Delta t\simeq T_{\!1\!/2}$. Blue elements are estimated from time-resolved intensity and phase correlation measurements directly. The gray elements represent the upper bound on the magnitude of the unknown off-diagonal elements, knowing the corresponding diagonal elements. The orange elements are estimated from the photon-number probabilities. The translucent region at the top of the blue bars indicate the standard deviation. (d) The density matrix computed using the theoretical model that includes emitter dephasing and pulses with a finite temporal width. For comparison to the experiment, the time bin threshold is chosen to correspond to the middle of the second pulse. Hence, in this case we have $p_{10}\simeq p_{01}$, unlike Eq. \ref{eq0011prbs} where $T$ is chosen to be just after the second pulse so that the probabilities take a simple analytic form. In panels (c) and (d), we only show half of the elements for visual clarity. The empty elements (white squares) are given by their symmetric counterpart. (e) Histogram of concurrence from sampling $10^5$ possible physical density matrices satisfying the partially reconstructed density matrix in panel (c) while including normally distributed measurement uncertainty (blue, broad peak) and without including any measurement uncertainty (red, narrow asymmetric peak). (f) Scatter plot of data composing panel (e). For visual clarity in panels (e) and (f), we only show 1/8 of the points sampled when neglecting measurement uncertainty.}
    \label{Sfigureconcurrencefidelity}
    \vspace{-3mm}
\end{figure*}

To supplement the two-pulse experiment at the half-life condition provided in the main text, we show self-homodyne measurements for 8 other pulse separations. However, instead of varying $T$ for each case, here we fix $T=\Delta t$.

For each $\Delta t$, we observe a clear quadratic signature in the phase-correlated data (see Fig.~\ref{Sfigure:rainbowfigure}(a)). In general, we find that the range of oscillation for both the self-homodyne signal $I_\text{SH}$ and intensity correlation $g^{(2)}_\text{HOM}$ increase with decreasing $\Delta t$. This trend is expected because both values quantify an intensity-normalised magnitude of coherence with the vacuum state. When rejecting coincident counts in the diagonal quadrants $ee$ and $ll$, we find that the overall value of $g^{(2)}_\text{HOM}$ decreases but the amplitude of its oscillation remains relatively unchanged, especially for larger $\Delta t$ (see Fig.~\ref{Sfigure:rainbowfigure}(b)). In this panel, for better visual demonstration, we have not re-normalised the counts. Hence, it corresponds to $g^{(2)}_\text{HOM,$el$}$ multiplied by the proportion $2\overline{\mu}_e\overline{\mu}_l$.

When we plot the scattered data in Fig.~\ref{Sfigure:rainbowfigure}(a) as a function of $\Delta t$, we find that the observed range of oscillation is very close to the ideal case (see Fig.~\ref{Sfigure:rainbowfigure}(c)). This theoretical range is obtained by first noting that, for an ideal $\ket{\phi^+}$ state, we have $\mu=2p_2$, $g^{(2)}=2p_2/\mu^2$, and $M=1/2$. Meanwhile, we can compute the squared magnitude of the second-order coherence ${c}^{(2)}=2p_0p_2/\mu^2$, where the factor of $2$ comes from the two-photon degeneracy. Hence, since $p_0=1-p_2$, we have that $2g^{(2)}_\text{HOM}=1-M+g^{(2)}-{c}^{(2)}\cos(2\phi)$ gives $M=1/2\leq g^{(2)}_\text{HOM}\leq 1/\mu=g^{(2)}$. Note that this observation is perfectly consistent with all states of the form $\ket{0}+\ket{2}$. Thus, to probe the Bell state specifically, it is necessary to subdivide the measurement into orthogonal time bins.

The necessary coherence measurements $M_{ee}$, $M_{ll}$, and ${c}^{(2)}_{el}$ needed to estimate the Bell-state fidelity are summarised in Fig.~\ref{Sfigure:rainbowfigure}(d) for each pulse separation. On the one hand, we notice that $M_{ee}$ and $M_{ll}$ approach the single-photon value of $M\simeq0.77$ when $\Delta t> T_{\!1\!/2}$, which confirms the convergence to the case of two sequential single photons $\ket{11}$. On the other hand, we can see a rise in ${c}^{(2)}_{el}$ as $\Delta t$ approaches zero, signifying the convergence to a state with significant vacuum component $\ket{00}$. The decrease in $M_{ee}$ and $M_{ll}$ here is due to the increased proportion of re-excitation noise to total emitted intensity and also due to the inability for the detector to resolve the short early photon. In between these extreme cases, we find the Bell state, which ideally would have $M_{ee}=M_{ll}={c}^{(2)}_{el}=1$.

\vspace{2mm}
With the measurements shown in Fig.~\ref{Sfigure:rainbowfigure}~(d) along with the photon-number probabilities in Fig.~\ref{Sfigure:pns}, we use Eq.~(\ref{supeq:phiplusfidelity}) to estimate the Bell-state fidelity (see Fig.~\ref{Sfigureconcurrencefidelity}~(a)). Following the approach detailed in section \ref{supsubsec:concurrenceestimates}, we also use the partially reconstructed density matrix to estimate the entanglement concurrence for each pulse separation (see Fig.~\ref{Sfigureconcurrencefidelity}~(b)). An example of the partially reconstructed density matrix for $\Delta t=98$ ps is shown in Fig.~\ref{Sfigureconcurrencefidelity}~(c) along with the corresponding theory prediction in Fig.~\ref{Sfigureconcurrencefidelity}~(d). Panels (e) and (f) of Fig.~\ref{Sfigureconcurrencefidelity} show the statistical distribution of concurrence from a sample of $10^5$ physical density matrices corresponding to the estimate of concurrence given in the main text. In these panels, we also show how the distribution would appear when neglecting the normally distributed measurement uncertainty (red, narrow asymmetric peak). By comparison to the broad normally distributed peak, we conclude that the variation in concurrence from the unmeasured density matrix elements is negligible compared to the variation due to the measurement uncertainty of the measured density matrix elements.

\vspace{2mm}
We find that our estimates of fidelity and concurrence peak for pulse separations around the half-life $\Delta t\simeq T_{\!1\!/2}$, as expected. The measured values also agree well with the general trend predicted by our theory and, in particular, agree that the maximum values are shifted slightly to shorter pulse separations due to dephasing \cite{scweinthesisSup}. In both cases, however, the model over-estimates the measured values because it does not take into account detector jitter, only finite pulse width and pure dephasing. Regardless, we see that the concurrence estimates are positive within the uncertainty of multiple standard deviations for all measured pulse separations, evidencing photon-number entanglement.

 \newpage
\bibliographystyle{naturemag}

\end{document}